\documentclass[conference]{IEEEtran}
\usepackage{caption,setspace,hyperref,color}
\usepackage{amsmath}
\usepackage[ruled, noend, noline]{algorithm2e}

\usepackage{enumitem}
\setlist{topsep=0pt, leftmargin=*}

\let\oldnl\nl
\newcommand{\nonl}{\renewcommand{\nl}{\let\nl\oldnl}}

\SetCommentSty{mycommfont}

\usepackage{amsthm}
\newcommand{\Def}[1]{\hyperref[def:#1]{Defn.\ref*{def:#1}}}

\newtheorem{problem}{Problem}

\usepackage{boldline, multirow}

\usepackage{graphicx}

\usepackage{subcaption}
\usepackage[font=small,labelfont=bf,
   justification=justified,
   format=plain]{caption}

\usepackage{xcolor}

\usepackage{bm}

\newcommand{\bi}{\textsc{Basic-Independent}}
\newcommand{\bd}{\textsc{Basic-Disjoint}}
\newcommand{\bc}{\textsc{Basic-Common}}
\newcommand{\mi}{\textsc{Max-Independent}}
\newcommand{\md}{\textsc{Max-Disjoint}}
\newcommand{\mc}{\textsc{Max-Common}}

\newcommand{\GD}{\textsc{Gene-Disease}}
\newcommand{\DB}{\textsc{DBpedia}}
\newcommand{\AT}{\textsc{Amazon-Tools}}
\newcommand{\AB}{\textsc{Amazon-Books}}

\newcommand{\YE}{\textsc{Yelp}}
\newcommand{\FO}{\textsc{Food}}

\newcommand{\GDx}{\textsc{GD}}
\newcommand{\DBx}{\textsc{DB}}
\newcommand{\YEx}{\textsc{YE}}
\newcommand{\ATx}{\textsc{AT}}
\newcommand{\ABx}{\textsc{AB}}

\newcommand{\FOx}{\textsc{FO}}

\newcommand{\floor}[1]{\left\lfloor #1 \right\rfloor}

\makeatletter
\newcommand{\newlineauthors}{%
  \end{@IEEEauthorhalign}\hfill\mbox{}\par
  \mbox{}\hfill\begin{@IEEEauthorhalign}
}
\makeatother

\def\BibTeX{{\rm B\kern-.05em{\sc i\kern-.025em b}\kern-.08em
    T\kern-.1667em\lower.7ex\hbox{E}\kern-.125emX}}

\begin{document}

\title{Retrieving Top-k Hyperedge Triplets: Models and Applications}

\author{\IEEEauthorblockN{Jason Niu\IEEEauthorrefmark{1}\IEEEauthorrefmark{2}}
\IEEEauthorblockA{
\textit{jasonniu@buffalo.edu}}
\and
\IEEEauthorblockN{Ilya D. Amburg\IEEEauthorrefmark{2}}
\IEEEauthorblockA{
\textit{ilya.amburg@pnnl.gov}}
\and
\IEEEauthorblockN{Sinan G. Aksoy\IEEEauthorrefmark{2}}
\IEEEauthorblockA{
\textit{sinan.aksoy@pnnl.gov}}
\and
\IEEEauthorblockN{Ahmet Erdem Sar{\i}y\"{u}ce\IEEEauthorrefmark{1}}
\IEEEauthorblockA{
\textit{erdem@buffalo.edu}}
\newlineauthors
\IEEEauthorblockA{\IEEEauthorrefmark{1}\textit{University at Buffalo}\\
    Buffalo, NY, USA}
\and
\IEEEauthorblockA{\IEEEauthorrefmark{2}\textit{Pacific Northwest National Laboratory}\\
Richland, WA, USA}
}

\maketitle

\begin{abstract}
Complex systems frequently exhibit multi-way, rather than pairwise, interactions. These group interactions cannot be faithfully modeled as collections of pairwise interactions using graphs and instead require hypergraphs. However, methods that analyze hypergraphs directly, rather than via lossy graph reductions, remain limited. Hypergraph motifs hold promise in this regard, as motif patterns serve as building blocks for larger group interactions which are inexpressible by graphs. Recent work has focused on categorizing and counting hypergraph motifs based on the existence of nodes in hyperedge intersection regions. Here, we argue that the relative sizes of hyperedge intersections within motifs contain varied and valuable information. We propose a suite of efficient algorithms for finding top-k triplets of hyperedges based on optimizing the sizes of these intersection patterns. This formulation uncovers interesting local patterns of interaction, finding hyperedge triplets that either (1) are the least similar with each other, (2) have the highest pairwise but not groupwise correlation, or (3) are the most similar with each other. We formalize this as a combinatorial optimization problem and design efficient algorithms based on filtering hyperedges. Our comprehensive experimental evaluation shows that the resulting hyperedge triplets yield insightful information on real-world hypergraphs. Our approach is also orders of magnitude faster than a naive baseline implementation.
\end{abstract}

% \begin{IEEEkeywords}
%     hypergraph, motif, combinatorial optimization, hyperedge triplet
% \end{IEEEkeywords}

\section{Introduction}
%Graphs are represented by a set of nodes and edges where each edge corresponds to a pairwise interaction.
Many complex systems contain higher-order interactions which are groupwise rather than pairwise.
These systems may suffer from significant information loss when modeled as a graph~\cite{zhou2006learning}.
For example, graph representations of author-paper~\cite{newman2001scientific}, company-board member~\cite{robins2004small}, and actor-movie~\cite{watts1998collective} networks link entities within one of these classes based on their group membership within the other. 
%(e.g. two authors are linked if they belong to the same paper).
Once such networks are modeled as graphs, groupwise relationships amongst entities cannot be distinguished from pairwise ones. %collection of nodes, it is impossible to distinguish the groupwise relationships this collection is a part of.
Finding dominant patterns within these groupwise relationships is an important part of categorizing today's web ecosystem.

Hypergraphs precisely capture these group relationships invisible in ordinary graphs.
A hypergraph consists of nodes and hyperedges, where each hyperedge can contain any number of nodes.
Hypergraphs can also be represented as a bipartite graph where nodes represent hypergraph nodes in one partition and hyperedges in the other;
bipartite edges link nodes across partitions if the corresponding hyperedge contains that node. 
%there exists an edge between a node and a hyperedge if the hyperedge contains the node.
Despite their broad applicability and expressivity, hypergraph algorithms are limited when compared to their graph counterparts. Furthermore, as shown in \cite{agarwal2006higher, hayashi2020hypergraph}, a number of existing hypergraph algorithms only consider pairwise (rather than groupwise) relationships, thereby effectively treating the hypergraph as a graph. Consequently, there is significant demand for efficient, hypergraph-{\it native} methods which utilize the higher-order relationships hypergraphs encode. 

%Therefore, there is significant demand for efficient algorithms which provide insightful information about these higher-order relationships.
One promising avenue for hypergraph-native analysis is motif discovery.
Motif discovery is effective in a myriad of graph mining tasks, such as controversy identification~\cite{coletto2017motif}, DNA analysis~\cite{ma2014motif}, and dense subgraph discovery~\cite{sariyuce2018peeling}.
In graphs, it is common for motifs to be based on triangles, which are cycles of length three and hence the smallest nontrivial dense subgraph.
While there is no consensus as to what constitutes a ``hypergraph triangle,'' a natural approach is to consider three hyperedges as the smallest nontrivial building blocks of hypergraph structure. Since hyperedges can contain more than two nodes, they can intersect in a wide variety of different ways that are impossible with three graph edges. 

Focusing precisely on these intersection patterns, recent work has aimed to classify and enumerate hypergraph motifs. 
%Recent work has focused on characterizing hypergraph motifs based precisely on these intersection relationships between hyperedges. 
Most relevant to our work, Lee et al.~\cite{lee2020hypergraph} define 30 $h$-motifs which exhaustively cover all ways three hyperedges can intersect, up to symmetries.
Crucially, these $h$-motifs are defined based on whether nodes {\it exist} within certain intersection regions between hyperedges. 
However, in practice, the {\it sizes} of intersection regions carry important information about the strength of overlap between hyperedge groups \cite{aksoy2020hypernetwork}. 
Consequently, $h$-motifs ignore key distinguishing qualities: the majority of motifs may belong to the same $h$-motif class and the intersection sizes within each motif may vary between being uniform and highly skewed.
This problem is exacerbated in the presence of large hyperedges, as high cardinality hyperedges increase the probability of having at least one node in intersection regions, thereby allowing a single type of $h$-motif's counts to dominate all the others.

For these reasons, we propose a quantitative approach to hypergraph motifs. We focus on sets of three hyperedges, or {\it hyperedge triplets}, and the sizes of their connectivity patterns. We call the nodes in one hyperedge but not the other two the triplet's {\it independent} regions, those in two but not the third the {\it disjoint} regions, and those in all three the {\it common} region.  
We formulate finding top hyperedge triplets as an optimization problem where we want to maximize the size of a specific region compared to the others.
This yields three related problems: finding hyperedge triplets that (1) are the least correlated with one another, (2) have the highest pairwise but not groupwise correlation, and (3) are the most correlated with one another as a group.
Our approach not only finds the global maxima for each of these three problems but also readily adapts to discover local maximum triplets---those maxima containing a given hyperedge---and can be extended to find larger subgraphs as well.
This local procedure can be used to give insight on behavioral patterns and the larger subgraphs formed from these triplets may represent larger, cohesive communities connected by a similar interest.
In this work, we focus on finding the top-$k$ hyperedge triplets where $k$ is a user-inputted positive integer.
Finding top-$k$ motifs has been previously studied with weighted triangles~\cite{kumar2020retrieving, taniguchi2022efficient, taniguchi2022efficient2, zhang2023top} and flow motifs~\cite{kosyfaki2019flow}.
%These problems correspond to the sizes of the independent regions, the disjoint regions, and the common region, respectively, of three hyperedges.
To the best of our knowledge, there is no prior work on finding size-aware motifs in hypergraphs.

To study this problem, we first consider a naive approach which either iterates over all triplets or cyclic-connected triplets of hyperedges.
We show that this approach has prohibitive time and space costs and is therefore not scalable for large hypergraphs. 
As a remedy, we propose a hyperedge avoidance scheme which skips over hyperedges based on their cardinalities and intersection sizes.
We also make use of a preprocessing routine which filters out inapplicable nodes and hyperedges and ranks hyperedges based on their cardinalities.
In an extensive experimental evaluation, we investigate the applicability of our algorithms on real-world networks through several case studies and application scenarios.
We also examine the runtime performance of our algorithms.

Our contributions are summarized as follows:
\begin{itemize}
    \item {\bf Novel problem formulation.} We formulate three maximization problems for finding top-$k$ hyperedge triplets which capture new relationships in hypergraph data. To the best of our knowledge, this is the first proposal for discovering the top-$k$ size-aware hypergraph motifs.
    \item {\bf Efficient algorithms.} We introduce algorithms for solving the three aforementioned maximization problems. Our algorithms, which iteratively update to-be-processed hyperedges based on the current maximum hyperedge triplet, improve upon a naive approach based on iterating over all candidate hyperedge triplets.
    %\item {\bf Finding maximum hyperedges.} We first consider a naive approach based on iterating over all candidate hyperedge triplets.
    %\item{\bf Improving the runtime.} We introduce a new approach which speeds up computation by iteratively updating to-be-processed hyperedges based on the current maximum hyperedge triplet.
    \item {\bf Experimental evaluation.} We evaluate our algorithms on several real-world hypergraphs to demonstrate practical runtimes. We also perform numerous case studies to illustrate the informative insights afforded by the top-$k$ hyperedge triplets in practice.
\end{itemize}

\section{Preliminaries} \label{sec:prelims} 
% A hypergraph is $H = (V, E)$ where $V$ is a set of elements called nodes and $E=(e_1,\dots,e_m)$ is an indexed family of sets, where $e_i \subseteq V$, called hyperedges.
% For $v \in V$, $E(v)$ is the set of hyperedges containing $v$, and $d_v=|E(v)|$ is the degree of $v$.
% For ease of notation, we let $n=|V|+|E|$ and $m=\sum_{v \in V} d_v$.
% Lastly, we use standard set notation: $X^c$ denotes the set of elements not in $X$, and ${X \choose k}$ denotes the set of $k$-element subsets of $X$.

A hypergraph is denoted as $H = (V, E)$ where $V$ is a set of elements called nodes and $E=(e_1,\dots,e_m)$ is an indexed family of sets where each $e_i \subseteq V$ is called an hyperedge.
For $v \in V$, $E(v)$ is the set of hyperedges containing $v$ and $|E(v)|$ is the degree of node $v$.
For each hyperedge $e_i \in E$, $i$ is the label for $e_i$ and $|e_i|$ is the size of $e_i$.
We call a set of three hyperedges $T=\{a,b,c\}$ a triplet.
Hyperedges within a triplet may be equivalent as sets but are always distinguishable by index.
We use ${X \choose k}$ to denote the set of $k$-element subsets of $X$.
Lastly, we define $n=|V|+|E|$ and $m=\sum_{v \in V} |E(v)|=\sum_{e \in E} |e|$.
\section{Related Work} \label{sec:related}

Here, we review prior work on bipartite and hypergraph motifs.

\noindent {\bf Butterflies.} Butterfly is a 2 x 2 biclique and represents the smallest unit of cohesion in bipartite graphs~\cite{wang2014rectangle, sanei2018butterfly}. 
There have been many works on butterfly counting~\cite{chiba1985arboricity, sanei2019fleet} along with efficient parallel approaches~\cite{shi2022parallel}.
Butterflies are commonly used as a basic motif for defining the community structure in bipartite networks, clustering coefficients~\cite{lind2005cycles, robins2004small}, and generative bipartite models~\cite{aksoy2017measuring}.
Sar{\i}y\"{u}ce and Pinar developed peeling algorithms based on butterflies for dense subgraph discovery
~\cite{sariyuce2018peeling}. 
%On a user-product network, they were able to identify groups of spammers which gave fake ratings without the use of any metadata.
The main drawback of the butterfly is its small size; it is restricted to two nodes and two hyperedges and has limited ability in capturing higher-order relations.
%In hypergraphs, however, butterflies limit the scope of hypergraph motifs due to its constraint on the number of nodes and hyperedges.

\noindent {\bf 6-cycles.} A 6-cycle is formed by two connected wedges (2-paths) which are closed by an additional wedge~\cite{yang2021efficient}.
% Thus, it can be thought of as a triangle of wedges whose projection to a graph forms a classic triangle.
6-cycles are effectively used as an alternative to butterflies in clustering coefficient definitions for bipartite graphs~\cite{schank2005approximating, opsahl2013triadic}.
One issue with a 6-cycle is that it does not account for the presence or lack of other possible edges among the 6 nodes.
As a remedy, an induced 6-cycle, which contains exactly 6 edges, is proposed as it forms a triangle in the unipartite projection with the minimal number of edges~\cite{niu2022counting}.
However, (induced) 6-cycles, like butterflies, offer a limited view on local structure as it has a restricted number of nodes and hyperedges.
%However, unlike motifs in bipartite graphs, hypergraph motifs often relate hyperedges together irrespective of the number of nodes.
% A significant problem of 6-cycles as an alternative to triangles in graphs is that it does not account for the number of edges, which may be up to 9 in the case of a 3 x 3 biclique.
% As a solution, induced 6-cycles may be considered because it forms a triangle in the projections with the minimal number of edges~\cite{niu2022counting}.

\noindent {\bf Higher-order motifs.} Proposed by Lotito et al.~\cite{lotito2022higher, lotito2023exact}, higher-order motifs provide an interesting approach for modeling higher-order relations in hypergraphs.
Given $k$ number of nodes, higher-order motifs are all the possible non-isomorphic connected hypergraphs (e.g., 6 motifs for $k=3$) where hyperedges can be overlapping. The authors provided a combinatorial characterization of higher-order motifs by giving upper and lower bounds for all possible motifs of nodes.
Significance profiles using motifs of size 3 and 4, with respect to the configuration model in~\cite{Chodrow20}, show that networks from the same domain exhibit similar trends. Higher-order motifs are based on a set number of nodes which can involve any number of hyperedges whereas our hyperedge triplets can consider any number of nodes but are limited to three hyperedges. In addition, we are interested in finding specific motif instances that optimize our measures rather than providing a global graph property.

\noindent {\bf $s$-walks.} Initially proposed by Aksoy et al. as a framework for hypergraph walks~\cite{aksoy2020hypernetwork}, $s$-walk is a sequence of hyperedges where each consecutive pair of hyperedges share at least $s$ nodes.
Aksoy et al. used $s$-walks as a basis for connected component analysis, closeness-centrality, and clustering coefficients in hypergraphs.
$s$-walks can be further classified as $s$-traces, $s$-meanders, and $s$-paths.
A closed $s$-walk of size three connects three hyperedges in a cyclic manner.
A closed $s$-walk is an $s$-trace if all hyperedges are unique as sets; an $s$-meander if it is an $s$-trace where no two hyperedge intersections are the same; and an $s$-path if it is an $s$-meander where no hyperedge intersection is a subset of another.

\begin{figure}[t]
    \centering
    \begin{subfigure}[h]{0.24\linewidth}
        \centering
        \includegraphics[width=\textwidth]{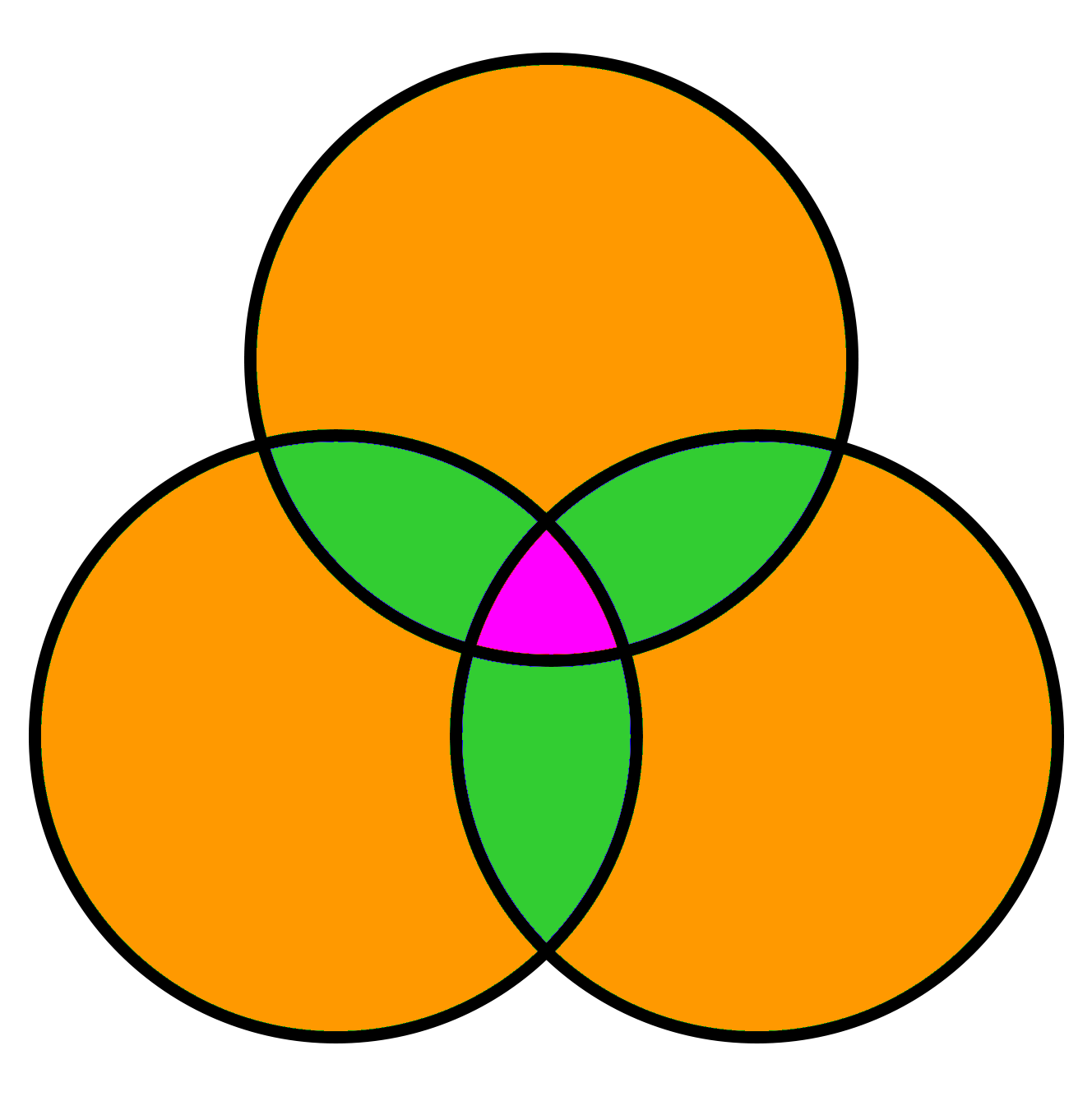}
        \caption{$h$-motif 4}
        \label{fig:4}
    \end{subfigure}
    \begin{subfigure}[h]{0.24\linewidth}
        \centering
        \includegraphics[width=\textwidth]{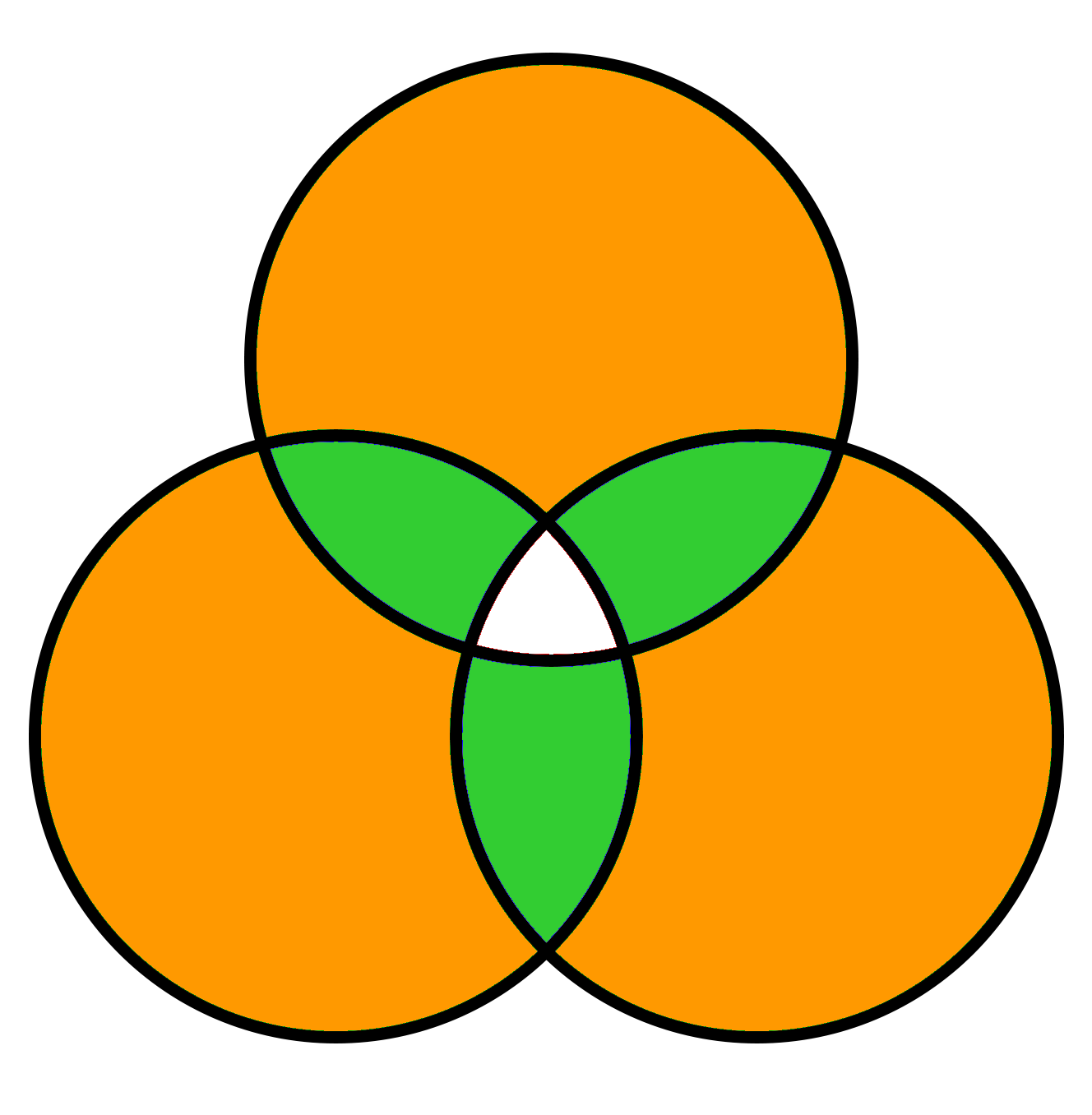}
        \caption{$h$-motif 8}
        \label{fig:8}
    \end{subfigure}
    \begin{subfigure}[h]{0.24\linewidth}
        \centering
        \includegraphics[width=\textwidth]{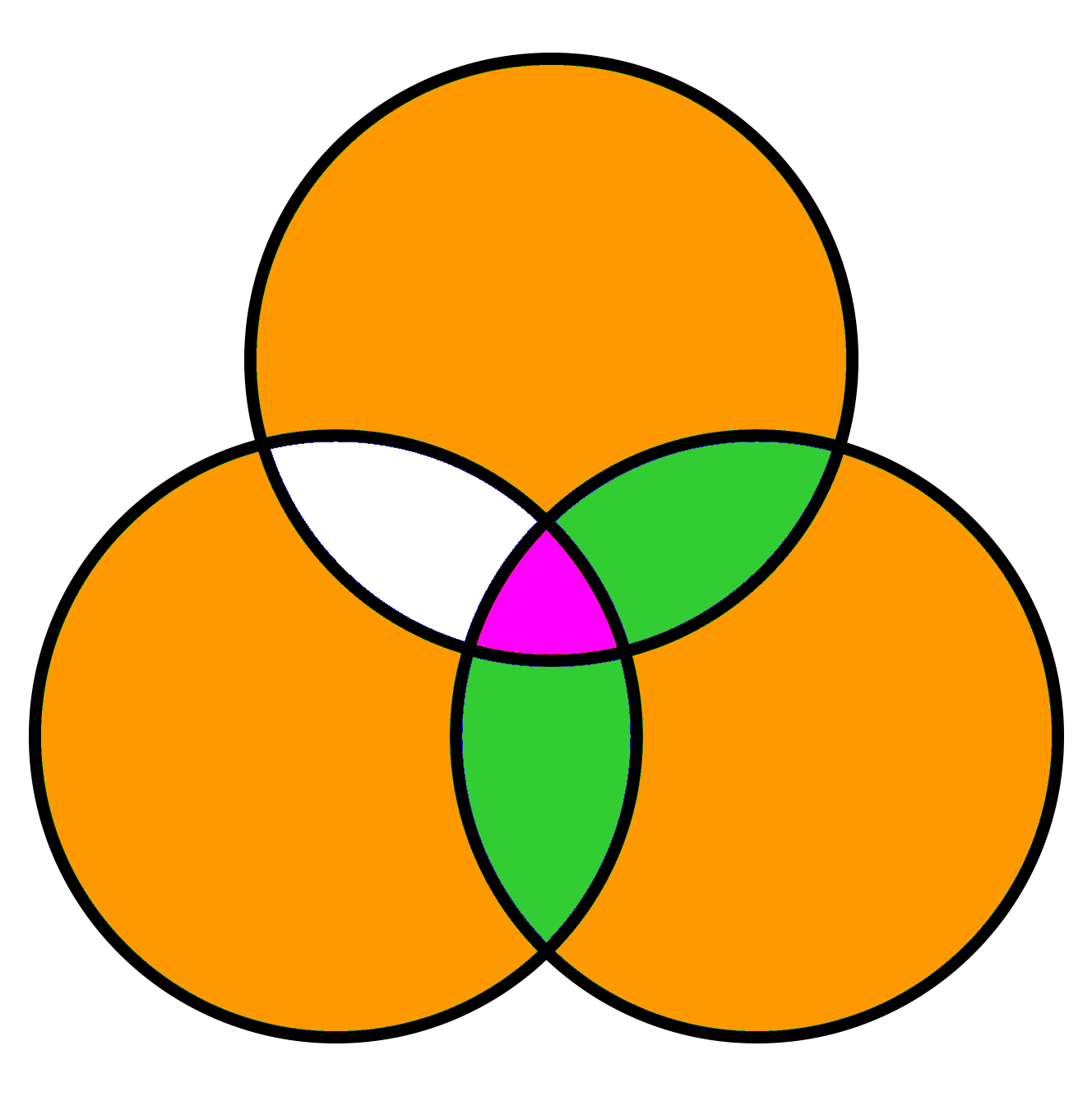}
        \caption{$h$-motif 14}
        \label{fig:14}
    \end{subfigure}
    \begin{subfigure}[h]{0.24\linewidth}
        \centering
        \includegraphics[width=\textwidth]{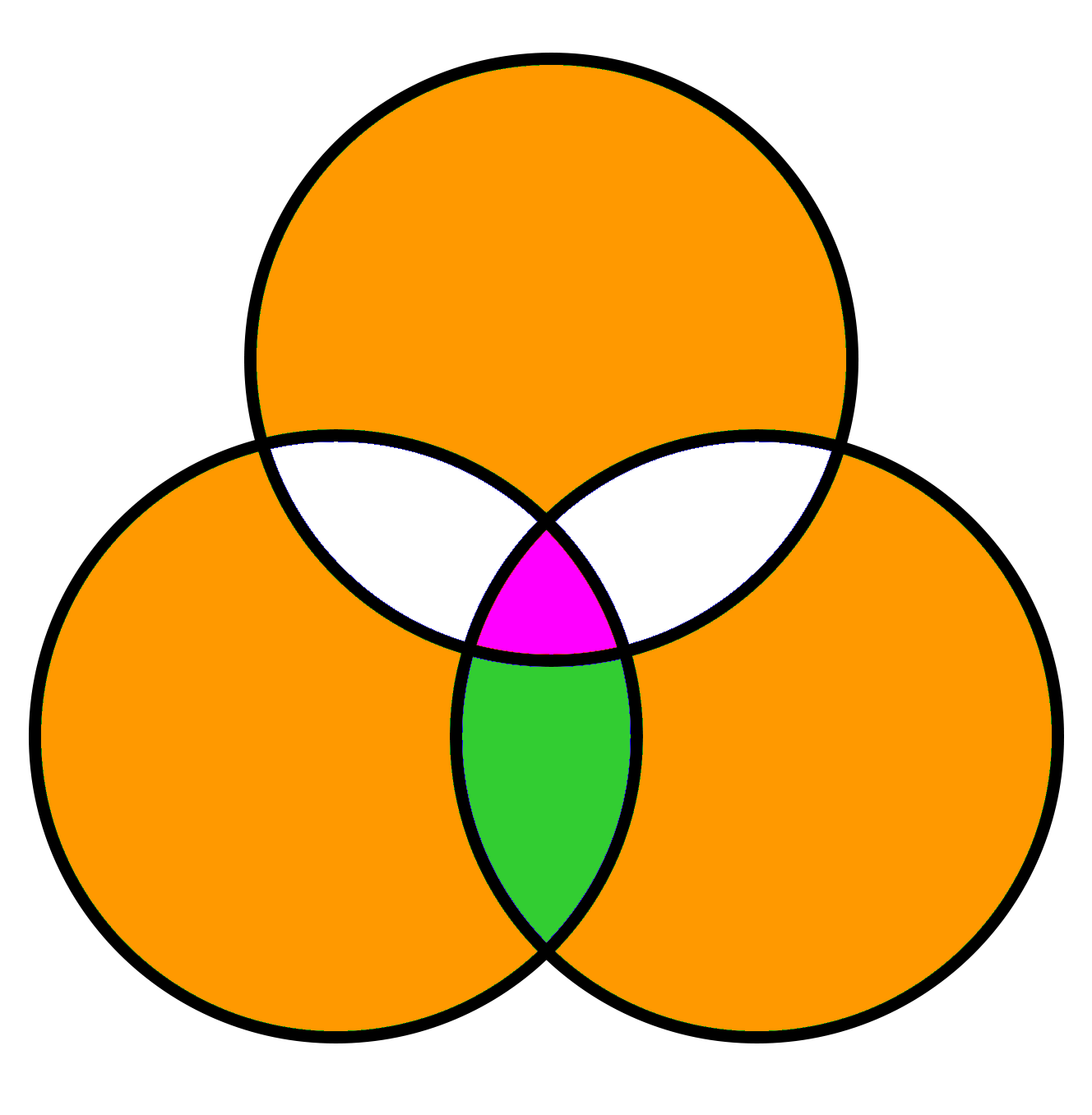}
        \caption{$h$-motif 20}
        \label{fig:20}
    \end{subfigure}
   \caption{A few examples of $h$-motifs, as denoted in~\cite{lee2020hypergraph}. Each circle denotes an hyperedge and intersections represent the set of common nodes. Colored-regions are non-empty.}
   \label{fig:hmotifs}
\end{figure}

\noindent {\bf $h$-motifs.} Lee et al. introduced $h$-motifs to describe the connectivity patterns of three connected hyperedges~\cite{lee2020hypergraph}.
Three hyperedges are defined to be connected if 1 of them is adjacent to 2 of the others.
Given a set of 3 connected hyperedges $\{e, f, g\}$, there exists 7 regions that describe the connectivity relations: $e \setminus (f \cup g)$, $f \setminus (e \cup g)$, $g \setminus (e \cup f)$, $(e \cap f) \setminus g$, $(f \cap g) \setminus e$, $(e \cap g) \setminus f$, and $e \cap f \cap g$ (see Figure~\ref{fig:hmotifs}).
Based on these regions, $h$-motifs describe all possible combinations of whether nodes exist in them, up to symmetry. There are 24 closed $h$-motifs where all three hyperedges are overlapping with each other. We show four high frequency closed $h$-motifs in Figure~\ref{fig:hmotifs}.
Lee et al. observed that $h$-motif distributions can be used effectively for evolution analysis of author-paper networks and hyperedge prediction models.
The aforementioned $s$-walk definition is related to $h$-motif in that an $h$-motif can be categorized more coarsely as a $1$-trace, $1$-meander, and/or $1$-path of length three.
We stress that $h$-motifs do not consider size information, unlike our work, as they only take into account the existence or lack of nodes in each region.
In a journal extension~\cite{lee2024hypergraph}, Lee et al. expanded upon $h$-motifs with $k$h-motifs by further classifying them into $k$ categories based on their region sizes $c$, e.g., 3h-motifs have three categories for each region: (1) c = 0, (2) c = 1, and (3) c $>$ 1.
This again has the same $h$-motif problem of larger hyperedges where all regions fall into category (3).
The authors motivate $h$-motifs and $k$h-motifs through the avenue of hypergraph classification, unlike our work which targets community detection.

%While characteristic profiles are used to motivate $h$-motifs, they are a global graph characteristic which do not apply to our work.
%However, not considering size allows for trivial $h$-motifs where the size of a region may be small compared to the other regions.

%{\bf Synthetic hypergraphs.} We will utilize real as well as synthetic hypergraph data for testing purposes. Synthetic graphs have traditionally been used when a real-world alternative is not readily available. For the creation of synthetic hypergraphs, it is desirable to reproduce the characteristics of real-world networks. Aksoy et al.~\cite{aksoy2017measuring} developed hypergraph Erd\H{o}s-R\'enyi and Chung-Lu implementations based on the work of Miller and Hagberg~\cite{miller2011efficient}.
%Since Erd\H{o}s-R\'enyi randomizes based on the probability of a vertex-hyperedge membership, it often generates hypergraphs of lower variance in terms of its degree distribution.
%Chung-Lu, however, tries to match the degree distribution of a real-world hypergraph, which keeps hyperedge cardinalities and vertex degrees in line with the original network.
%In this work, we generate synthetic hypergraphs with Aksoy et al.'s Erd\H{o}s-R\'enyi and Chung-Lu adaptations.
\section{Motivation}
Before formalizing our problem statements, we first go over the motivation in light of our preceding discussion of related work.
%go over the motivation behind our problem statements which we formalize in Problem\ \ref{prob:objective}.

The motivation behind Lee et al's $h$-motif~\cite{lee2020hypergraph} and Aksoy et al.'s $s$-walk~\cite{aksoy2020hypernetwork} works (and by extension, ours) is in the analysis of the structure of hypergraphs.
These higher-order methods operate directly on the hyperedges and reveal structures that cannot be detected by graph-based measures.
Our hyperedge triplet approach is more related to $h$-motifs~\cite{lee2020hypergraph} and $s$-walk~\cite{aksoy2020hypernetwork} based motifs than bipartite cycle-based motifs~\cite{yang2021efficient, niu2022counting} or higher-order motifs~\cite{lotito2022higher}.
Butterflies and 6-cycles constrain the number of hyperedges {\it and} nodes in a motif to two or three, whereas our study considers three hyperedges containing any number of nodes.
Like our approach, $h$-motifs also contain exactly three hyperedges and an unconstrained number of nodes.
Since the intersection relationships in $h$-motifs are more refined than those in $s$-walks, we focus on motivating our method in comparison to the existing $h$-motif approach.
However, note that the authors in~\cite{lee2020hypergraph} use $h$-motifs in hypergraph classification (of small hypergraphs) while our work is based on community detection (where the hypergraph may be large).

\begin{figure}[!t]
    \centering
    \includegraphics[width=0.7\linewidth]{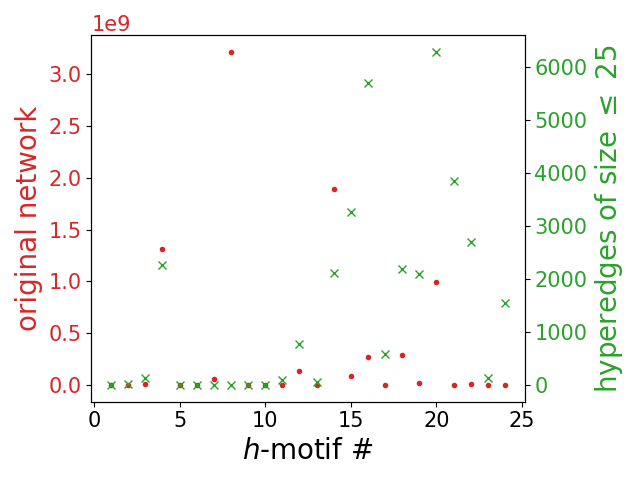}
    \caption{Closed $h$-motif counts for\ \GD. The x-axis contains arbitrary motif numberings for the closed $h$-motifs in~\cite{lee2020hypergraph}. Figure \ref{fig:hmotifs} shows the top four $h$-motifs in the original network.}
    \label{fig:GD_hcnts}
\end{figure}

%\erdem{following is revised}
Unlike $h$-motifs which analyze all structures, of which many are trivial, our work provides a framework for finding the best structures and relationships in a hypergraph.
The key difference between our work and $h$-motifs is that we explicitly take the sizes of intersecting regions into account.
As we are the first work to find these interesting structures using the intersection sizes of hyperedges, our work discovers new insights unable to be uncovered by previous methods.
Factoring in the sizes of intersections has strong ramifications when applying these methods to real-world hypergraphs.
For example, it is noteworthy that when testing the applicability of $h$-motifs on real-world datasets, Lee et al.~\cite{lee2020hypergraph} only tested on hypergraphs which filter their hypergraph data to keep the maximum hyperedge size as 25.
%\erdem{did u check that work? are they really filter large hyperedges and justify it some way?}
%\jason{reworded; they just used only datasets which filtered out all cardinalities > 25, they didnt actually filter it out themselves. they gave no justification.}
While in some contexts removing hyperedges of certain cardinalities might be justifiable~\cite{landry2023filtering} or part of data cleaning, ignoring large hyperedges does lead to information loss, especially in contexts where large hyperedges are prevalent such as diseases that connect thousands of genes or products with a large number of reviews.
To make it concrete, we offer a preliminary study on how such filterings impact the occurrence of $h$-motifs.
%We consider the raw unfiltered \GD~\cite{pinero2016disgenet} dataset (see Table \ref{tab:datasets}).
We give the $h$-motif distribution for the raw unfiltered \GD~\cite{pinero2016disgenet} dataset (see Table \ref{tab:datasets}) against a filtered \GD\ with hyperedges of cardinality at most 25 in Figure~\ref{fig:GD_hcnts}.
Removal of higher cardinality hyperedges causes the $h$-motif distribution to be relatively flat compared to the original version---note that motif counts reduce from billions to thousands.
Furthermore, in the unfiltered version, it is striking that four specific $h$-motif counts dominate the whole distribution---namely motifs 4, 8, 14, and 20 where at least five of seven regions are non-empty (shown in Figure~\ref{fig:hmotifs}).
In our evaluation, we found this occurrence to be common in real-world hypergraphs when there are a significant number of high cardinality hyperedges.
%Note that those four motifs have non-empty counts in most regions.
This suggests that larger hyperedges---a common occurrence guaranteed by the heavy-tailed degree distributions---may skew the $h$-motif distributions, posing a challenge for the expressivity of the $h$-motif framework.

To investigate this further, we generate synthetic versions of \GD\ and examined the $h$-motif distribution.
We use the Erd\H{o}s-R\'enyi and Chung-Lu hypergraph models developed by Aksoy et al.~\cite{aksoy2017measuring} based on the work of Miller and Hagberg~\cite{miller2011efficient}.
Since Erd\H{o}s-R\'enyi randomizes based on the probability of a vertex-hyperedge membership, it often generates hypergraphs of lower variance in terms of its degree distribution.
Chung-Lu, however, tries to match the degree distribution of a real-world hypergraph, which keeps hyperedge cardinalities and vertex degrees in line with the original network.
%Since the Erd\H{o}s-R\'enyi model does not consider the degree distribution of real-world hypergraphs, it often results in hypergraphs with low variance in their hyperedge cardinalities.
Results suggest that the Erd\H{o}s-R\'enyi version typically does not have any significant disparity in its $h$-motif counts, unlike the original network and similar to the results from filtering high cardinality hyperedges.
However, the Chung-Lu model, which matches the degree distribution in expectation, had a similar $h$-motif distribution compared to the original \GD\ network.
The reasoning for this is clear: higher cardinality hyperedges often participate in more $h$-motifs compared to hyperedges with lower cardinalities.
Due to the large number of nodes, these $h$-motifs also frequently have at least one node in each of their regions.
Thus, the evaluation on synthetic data also suggests that the large hyperedges may obscure and confound $h$-motif distribution patterns.

For our model, we work on hyperedge triplets ranked by the custom weights (function of sizes) of their intersection patterns.
This is similar to the concept of weighted triangle in edge-weighted graphs~\cite{czumaj2009finding}.
Weighted triangles are 3-cliques which are commonly measured by either the mean or the sum of the triangle's edge weights.
Benson et al. studied using weighted triangles for higher-order link prediction where the goal is to predict which groups of nodes are most likely form a new simplex (a group interaction)~\cite{benson2018simplicial}.
As such, these simplexes essentially represent new hyperedges.
Their analysis suggests that triples of nodes with strong ties are most likely to form new simplexes in the future.
Inspired by Benson et al.'s work, Kumar et al. proposed fast algorithms for finding the top-$k$ weighted triangles~\cite{kumar2020retrieving}.
There are also many other works which study top-$k$ weighted triangles in recent years~\cite{taniguchi2022efficient, taniguchi2022efficient2, zhang2023top}.
In our work, we find the top-$k$ hyperedge triplets which adapts the concept of a weighted triangle for hypergraphs.
%\erdem{i'm unsure about putting the following sentence as it may trigger some reviewer to say why we haven't evaluated hyperedge prediction. It's a nice thought though, let's keep that in mind. Let me know if you disagree}
%\jason{Moved to conclusion and future work section}
%These hyperedge triplets can then be extended for hyperedge prediction algorithms as those with the strongest ties (or weights) are more likely to involve nodes in new hyperedges.

\section{Problem Statement} \label{sec:prob}

\begin{figure}[!b]
    \centering
    \begin{subfigure}[h]{0.49\linewidth}
        \centering
        \includegraphics[width=\textwidth]{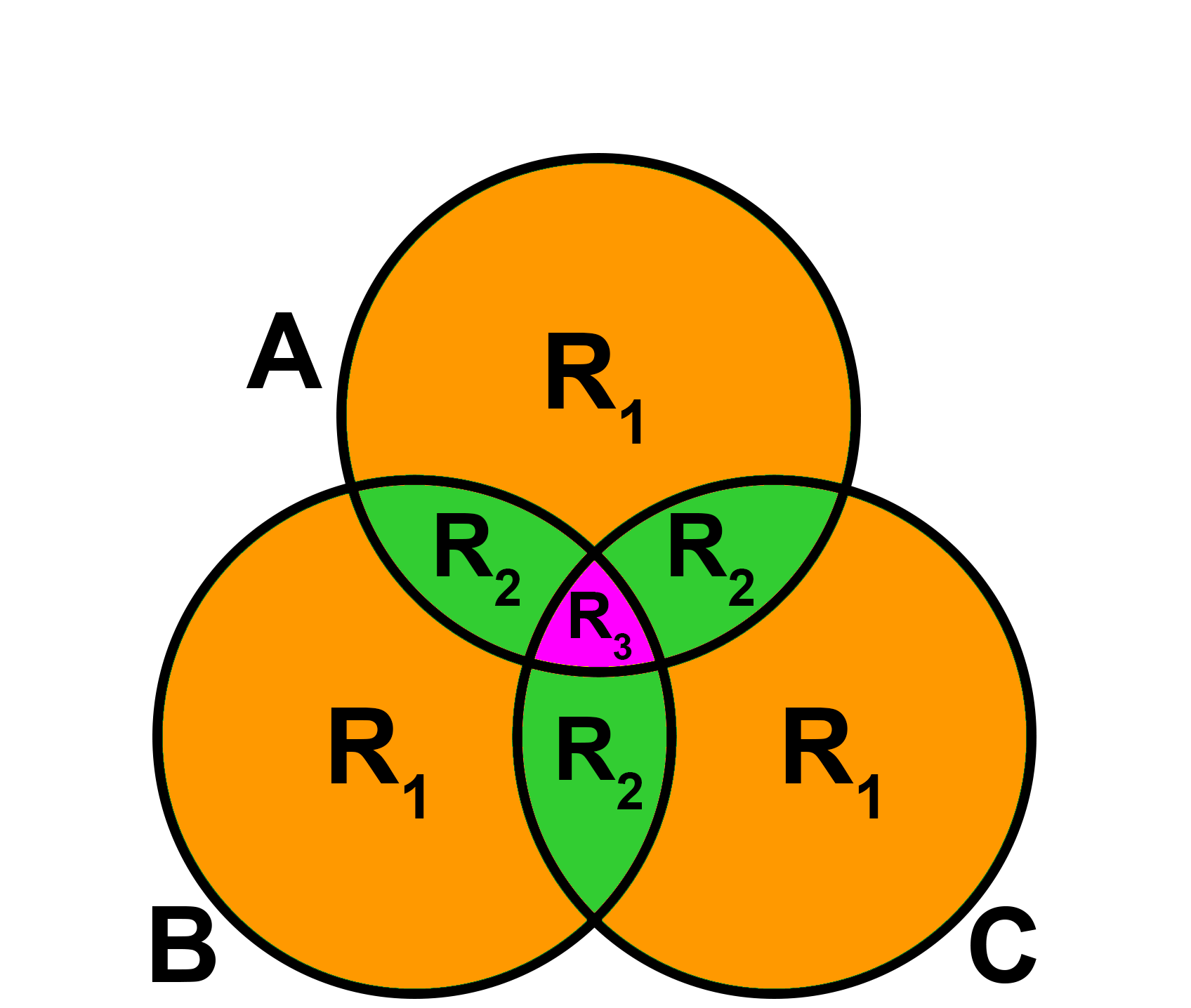}
        \caption{Region labels.}
        \label{fig:regionLabels}
    \end{subfigure}
    \begin{subfigure}[h]{0.49\linewidth}
        \centering
        \includegraphics[width=\textwidth]{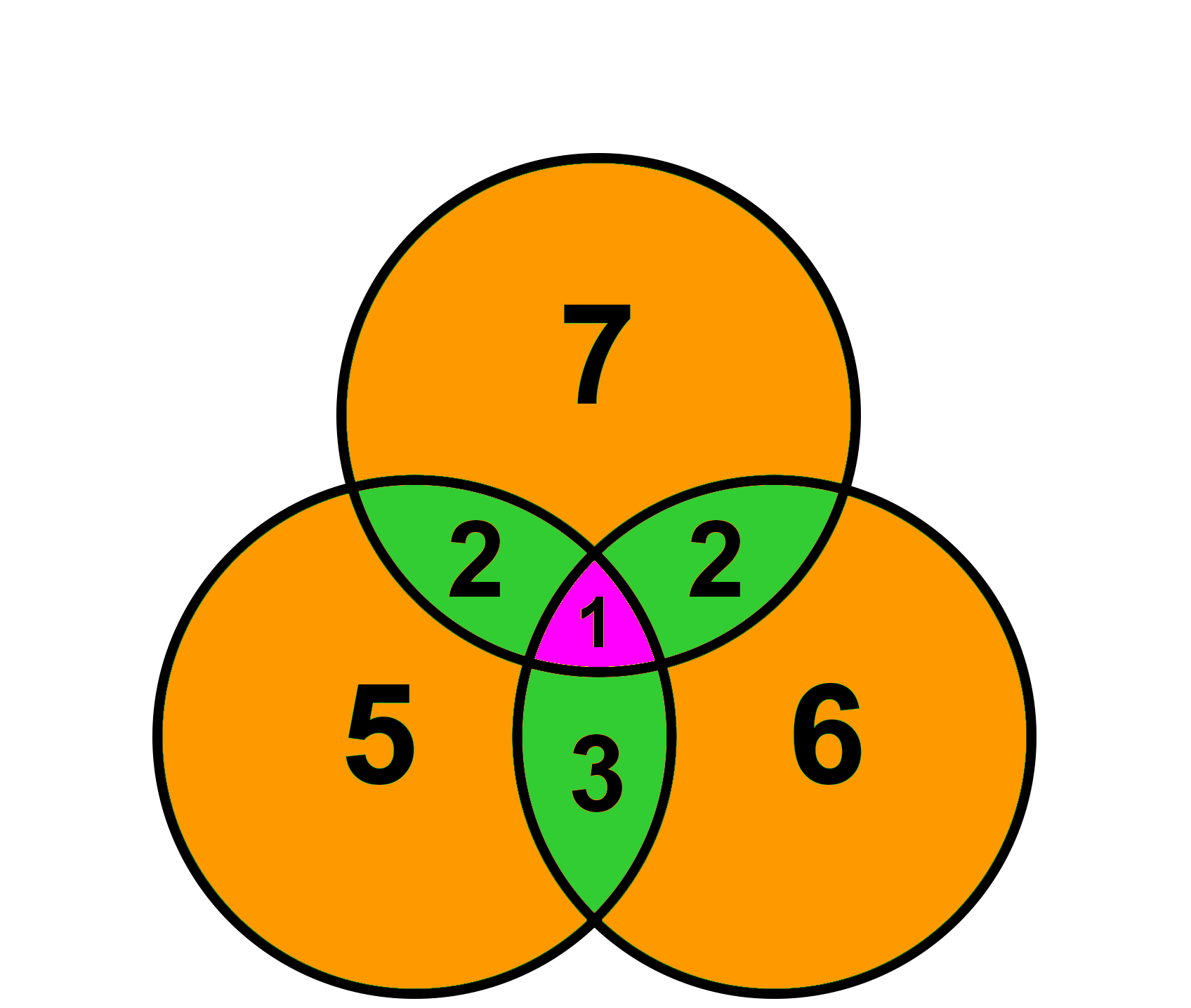}
        \caption{A toy triplet.}
        \label{fig:weightExample}
    \end{subfigure}
   \caption{Hyperedge triplet regions. Figure \ref{fig:regionLabels} depicts the independent ($R_1$), disjoint ($R_2$), and common ($R_3$) regions for the hyperedge triplet $\{A, B, C\}$. Figure \ref{fig:weightExample} shows a toy triplet where the independent weight is $W_1(T)=\frac{\min(7, 5, 6)}{1+(2+2+3+1)}=\frac{5}{9}$, the disjoint weight is $W_2(T)=\frac{\min(2, 2, 3)}{1+1}=1$, and the common weight is $W_3(T)=\frac{\min(1)}{1}=1$.}
   \label{fig:regions}
\end{figure}

Our approach focuses on the sizes of the different ``region types'' defined by the intersection relations within a triplet. To define these regions formally, let $X\subseteq T$ denote a subset of hyperedges within a triplet $T$. Then $N:=N_T$ is defined by
\[
N(X)=\left(\bigcap_{x \in X} x \right) \setminus \left(\bigcup_{x \in T \setminus X} x \right).
\]
Put equivalently, $N(X)$ picks out the elements shared across every hyperedge within $X$ that are not in other hyperedges (in $T$). Applying this function to all subsets of $T$ yields a partition, as visualized in Figure \ref{fig:regionLabels}.
These regions fall into three categories which we call the {\bf independent ($R_1$), disjoint ($R_2$)}, and {\bf common ($R_3$)} regions:

\begin{itemize}
    \item {\bf $R_1$} = $\{N(a), N(b), N(c)\}$
    \item {\bf $R_2$} = $\{N(a,b), N(a,c), N(b,c)\}$
    \item {\bf $R_3$} = $\{N(a,b,c)\}$
\end{itemize}

\noindent %We denote independent as $R_1$ to represent 1-way relationships, disjoint as $R_2$ for 2-way relationships, and common as $R_3$ for 3-way relationships.
We subsequently use these 3 region types to formulate 3 notions of hyperedge triplet {\it weights}. Namely, for a hyperedge triplet $T$, its independent $W_1$, disjoint $W_2$, and common weight $W_3$ is given by:
\begin{equation} \label{eq:wght}
W_j(T)=\frac{\min_{X \in {T \choose j}}|N(X)|}{1+\sum_{i>j}\sum_{X \in {T \choose i}} |N(X)| }\\
\end{equation}
\noindent We take the minimum size of the target regions in the numerator to prevent the misrepresentation of hyperedge triplets if one region dominates the others.
This is based on the concept that a group is only as strong as its weakest link.
We force all three hyperedges to have high cardinalities which leads to insightful relationships, unlike other aggregation functions such as sum and mean.
To avoid an undefined value when all antagonistic regions are empty, we increment the denominator by one.
Division is used to reward a significant proportion of nodes in the desired regions with the possibility of a few outliers.
Each weight definition ($W_1$, $W_2$, $W_3$) considers only the regions involving the number of hyperedges in their subscript.
For $W_1$, we consider any region involving any single hyperedge, which is all 7 regions.
For $W_2$, we consider any region involving any two hyperedges, which includes both $R_2$ and $R_3$.
For $W_3$, we consider any region involving all three hyperedges, which is only $R_3$.
%\erdem{we still don't have an explanation for why the denominator is different for different weights, remember there was a reviewer comment for that in webconf. Relatedly, we must explain why the independent is R1, disjoint is R2 and common is R3 (i.e., why we order them this way).}
Lastly, while we focus on triplets in this work, we note the above definition may be applied to $k$-tuples of hyperedges for $k>3$, yielding $k$-many different region types. 
Figure~\ref{fig:weightExample} gives a toy example where $W_1(T)=\frac{\min(7, 5, 6)}{1+(2+2+3+1)}=\frac{5}{9}$, $W_2(T)=\frac{\min(2, 2, 3)}{1+1}=1$, and $W_3(T)=\frac{\min(1)}{1}=1$.

To account for the role that size plays in hypergraph motifs, we consider the independent, disjoint and common weights to define ``good'' hyperedge triplets. 
As we do not constrain the number of nodes in a given region, we focus not on enumerating motifs with a given intersection region size distribution but rather on finding the ``optimal'' motifs with regard to their size patterns.
We thus cast this as a maximization problem where we seek hyperedge triplets with the highest weights in Equation~\ref{eq:wght}.
Our weight definition ensures that in the optimal triplets a significant proportion of nodes are evenly distributed across the corresponding region types.
Our problem is defined as follows:
% To design a more definitive model for $h$-motifs, we can consider the sizes of each independent and overlapping region.
% However, there can be an infinite number of combinations of sizes of three connected hyperedges.
% Instead of finding all combinations, we can focus on the most important combinations.
% Therefore, we can design this as a maximization problem where we want to find the combinations which involve the maximum sizes.
% We can also expand this problem to include all sets of three unique hyperedges regardless of their connectivity (a hyperedge triplet).
 %, as supported in Section \ref{sec:entropy}.
\begin{problem}\label{prob:objective}
    For a hypergraph with hyperedge set $E$, the maximum $\{\mbox{independent, disjoint, common}\}$ problem is to find the $T \in {E \choose 3}$ that maximizes $\{W_1(T), W_2(T), W_3(T)\}$.
\end{problem}
\noindent In our implementation, we find the top-$k$ hyperedge triplets where $k$ is a user-inputted positive integer.
These hyperedge triplets can then be used as building blocks for larger connected structures.
%\noindent For Problem \ref{prob:objective}, if we simply find the hyperedge triplet corresponding to the largest region-of-interest size, we may obtain a triplet where the size may be small compared to the total size of all unwanted regions.
%Therefore, we factor in the size of all unwanted regions in our problem statements to avoid the misrepresentation of hyperedge triplets.\sinan{Jason does this paragraph still fit here? Doesn't make sense}

\noindent {\bf Remark.} We want to emphasize that a few other alternative, and intuitive, weight definitions have significant flaws that may lead to misleading results.
One such formulation would be the proportion of nodes in a specific region compared to all regions.
However, such an objective function would be misleading as the weight definition would be normalized and unable to adequately address the cardinalities of hyperedges---a triplet with millions of nodes could be ranked the same as a trivial triplet with only a few nodes.
Another formulation that divides by the sum of all $other$ regions would be too strict in sparse hypergraphs where the sizes in $R_1$ frequently dominates those in $R_2$ and $R_3$.
This would result in only a few, if any, significant disjoint and common hyperedge triplets.
Using $min()$ in the denominator would also lead to misleading results as three hyperedges may be labeled as independent even though two of the three have significant overlap.
Other aggregation functions such as $sum()$ and $mean()$ in the numerator may result in an empty hyperedge participating in a triplet with two large hyperedges, which is a trivial relationship.
Our formulation involves a layered approach with $R_1$ $\rightarrow$ $R_2$ $\rightarrow$ $R_3$ where we take the minimum size of all regions in the current layer and divide by the sum of the sizes of all deeper layers.
We also believe that this formulation can naturally be extended to any number of hyperedges.
For example, with four hyperedges, we can incorporate another layer $R_4$ for four-way relationships which results in an additional layer after $R_3$.

\section{Algorithms} \label{sec:alg}
We start by introducing a naive approach in Section~\ref{sec:naiveAlgs} which iterates over the set of candidate hyperedge triplets for each maximization problem.
Then, we propose a new technique which filters future hyperedge triplets based on the previously visited maximum ones in Section~\ref{sec:maxAlgs}.
Lastly, we show how our algorithms can be adapted for local search and introduce a merging approach to find larger subgraphs of triplets in Section~\ref{sec:app}.

\subsection{Enumerating through Candidate Sets} \label{sec:naiveAlgs}
Here, we introduce a common framework for the three baseline algorithms in Algorithm~\ref{alg:basic}, {\sc Basic}.
It has three variants---{\sc Independent, Disjoint, Common}---and all scan through all the triplets in their candidate sets.
The candidate set for each problem is the minimal set of hyperedge triplets that contains the desired maximum triplet.

\setlength{\textfloatsep}{0pt}
\begin{algorithm}[!t]
\caption{\textsc{Basic}~($H$, $k$, $\alpha$)}
\label{alg:basic}
\DontPrintSemicolon
  {\bf Input:} $H = (V, E)$: hypergraph, $k \in [1, \infty)$, $\alpha \in \{1, 2, 3\}$: algorithm type, corresponding to {\sc Independent, Disjoint, Common}, respectively \;
  {\bf Output:} $\triangle_k$: top-$k$ hyperedge triplets\;
    \nl $\triangle_k \gets \emptyset$; $\triangle_w \gets 0$\;
    \nl rename hyperedges in $E$ by decreasing degrees\; \label{line:bi:rename}
    \nl sort neighbor lists in $H$ in ascending order\; \label{line:bi:sort}
    \nl $S \gets \emptyset$ \tcp*{hashmap of hyperedge pairs to node sets} \label{line:bi:S}
    \nl \ForEach {$e_x \in E$}{ \label{line:bi:aStart}
        \nl \ForEach {$u \in e_x$}{
            \nl \ForEach {$e_y \in E(u)$ s.t. $y > x$}{
                \nl $S[x, y] \gets S[x, y] \cup u$\;
            }
        }
    }  \label{line:bi:aEnd}
    \nl \tcp{$C(\alpha)$ is candidate set; which is all triplets for {\sc Independent} and closed triplets for {\sc Disjoint/Common}}
    \ForEach {$e_x, e_y, e_z \in C(\alpha)$ s.t. $x < y < z$ and $|e_z| > \triangle_w$} { \label{line:bi:tripletStart}
        \nl $w \gets$ weight of $\{e_x, e_y, e_z\}$\; \label{line:bi:weight}
        \nl \If {$w > \triangle_w$}{
            \nl {\bf if} $|\triangle_k| < k$ {\bf then} $\triangle \gets \triangle \cup \{e_x, e_y, e_z\}$\;
            \nl \If {$|\triangle_k| = k$}{
                \nl $\triangle_{min} \gets $ minimum weight triplet in $\triangle_k$\;
                \nl replace $\triangle_{min}$ in $\triangle_k$ with $\{e_x, e_y, e_z\}$\;
                \nl $\triangle_w \gets$ minimum weight in $\triangle_k$\;
            }
        }
    } \label{line:bi:tripletEnd}
    \nl \textbf{return} $\triangle_k$
\end{algorithm}

% For MIWP (Problem\ \ref{prob:int}), the ideal maximum hyperedge triplet contains three hyperedges with large cardinalities which do not have any overlap with each other.
% However, in real-world networks, this no overlap constraint is often too strict, resulting in lower quality triplets.
% As a solution, we can allow for some nodes to be in multiple hyperedges if there are a significant number of nodes not in any overlapping region.
% Therefore, we need to consider all hyperedge triplets in the candidate set for MIWP.

The input is a hypergraph $H = (V, E)$, a positive integer $k$, and an algorithm type $\alpha$, which is equal to 1, 2, or 3, for {\sc Independent, Disjoint,} and {\sc Common} variants, respectively.
We begin by modifying hyperedge ids and sorting neighbor lists such that hyperedges with lower ids have higher cardinalities than those with higher ids (lines \ref{line:bi:rename} and \ref{line:bi:sort}).
By organizing hyperedges based on their cardinalities, we can process higher cardinality hyperedges first, increasing the chance of finding the maximum hyperedge triplet at an earlier iteration.
% The sorted neighbor lists allow for efficient enumeration of hyperedges with higher ids and common region computations.
Afterwards, we store all non-empty pairwise intersections between two hyperedges in a container $S$ (line \ref{line:bi:S}).
Then, we iterate through all candidate hyperedge triplets where the lowest cardinality hyperedge is greater than the current maximum weight (line \ref{line:bi:tripletStart}).
The candidate set for the {\sc independent} variant ($C(1)$) is simply all $\binom{E}{3}$ hyperedge triplets.
For the other variants, the candidate set ($C(2)$ and $C(3)$) is all the closed hyperedge triplets: $e_x, e_y, e_z \in \binom{E}{3}$ s.t. $|e_a \cap e_b| > 0$ $\forall$ $e_a,e_b \in \binom{\{e_x, e_y, e_z\}}{2}$. We find closed hyperedge triplets in $O(1)$ time by performing a lookup on $S$ for overlapping hyperedges.
To speed up the weight computation on line \ref{line:bi:weight}, we use the pairwise intersections stored in $S$ to limit duplicate common neighbor operations.
Finally, the hyperedge triplets with the highest weights are returned.

% {\it Time Complexity.} Lines \ref{line:bi:aStart}-\ref{line:bi:aEnd}\ traverses through all paths of length 2 which contain two hyperedges and takes $O(|V| \cdot {\langle d_V \rangle} \cdot {\langle d_U \rangle}) = O(m \cdot {\langle d_U \rangle})$ time.
% Lines \ref{line:bi:tripletStart}-\ref{line:bi:tripletEnd} computes the independent weight for all hyperedge triplets and takes $O({|V| \choose 3} \cdot {\langle d_V \rangle})$ time, which is equivalent to the total time complexity of \bi.

{\it Time Complexity.} Lines \ref{line:bi:aStart}-\ref{line:bi:aEnd}\ traverses through all paths of length 2 which contain two hyperedges and takes $O(m \cdot |E|)$ time.
Lines \ref{line:bi:tripletStart}-\ref{line:bi:tripletEnd} compute the weight for all hyperedge triplets and takes $O({|E| \choose 3} \cdot |V|)$ time, which is equivalent to the total time complexity.

% {\it Space Complexity.} In addition to the $O(m)$ space required for the graph, we store the intersection between every pair of intersecting hyperedges in a container $S$ (Line \ref{line:bi:S}). Therefore, the space complexity is at most $O(|V| \cdot  {\langle d2_V \rangle} \cdot {\langle d_V \rangle}) = O(m \cdot {\langle d2_V \rangle})$. Since real-world hypergraphs are typically very sparse, the required space is usually much less than this maximum.

{\it Space Complexity.} In addition to the $O(m)$ space required for the graph, we store the intersection between every pair of intersecting hyperedges in a container $S$ (line~\ref{line:bi:S}). Therefore, the space complexity is at most $O(m \cdot |E|)$. Since real-world hypergraphs are typically very sparse, the required space is usually much less than this maximum.

\subsection{Avoiding Irrelevant Triplets} \label{sec:maxAlgs}
In this section, we introduce our main algorithms, \mi, \md, and \mc.
These algorithms skip the processing of irrelevant hyperedges with low cardinalities.

\setlength{\textfloatsep}{0pt}
\begin{algorithm}[!t]
\caption{\textsc{Max}~($H$, $k$, $\alpha$)}
\label{alg:max}
\DontPrintSemicolon
  {\bf Input:} $H = (V, E)$: hypergraph, $k \in [1, \infty)$, $\alpha \in \{2, 3\}$: algorithm type, corresponding to {\sc Disjoint, Common}, resp.\;
  {\bf Output:} $\triangle_k$: top-$k$ hyperedge triplets\;
    \nl $\triangle_k \gets \emptyset$; $\triangle_w \gets 0$\;
    \nl rename hyperedges in $E$ by decreasing degrees\; \label{line:m:rename}
    \nl sort neighbor lists in $H$ in ascending order\; \label{line:m:sort}
    \nl $S \gets \emptyset$ \tcp*{hashmap of hyperedge pairs to node sets} \label{line:m:s}
    \nl \tcp{bool. expression depending on alg. type}
    $\theta(e, \triangle_w, \alpha) \gets \floor{\frac{|e|}{2}} > \triangle_w$ {\bf if} $\alpha=2$ {\bf else} $|e| > \triangle_w$\;
    \nl \ForEach {$e_y \in E$ s.t. $\theta(e_y, \triangle_w, \alpha)$}{ \label{line:m:b}
        \nl $T \gets \emptyset$ \tcp*{hashmap of hyperedges to node sets} \label{line:m:t}
        \nl \ForEach {$u \in e_y$}{
            \nl \ForEach {$e_z \in E(u)$ s.t. $z > y$ and $\theta(e_z, \triangle_w, \alpha)$}{ \label{line:m:c}
                \nl $T[z] \gets T[z] \cup u$\;
            }
        }
        \nl \ForEach {$z, N_{yz} \in T$ s.t. $|N_{yz}| > \triangle_w$}{
            \nl \tcp{skip pair based on pairwise upper bound}
            {\bf if} $|N_{yz}| > \triangle_w$ {\bf then} $continue$\;
            \nl \ForEach {$x, N_{xy} \in S[y]$ s.t. $x \in S[z]$}{
                \nl $v \gets min(|N_{xy}|, |S[z][x]|, |N_{yz}|)$\;
                \nl \tcp{skip common region calculation based on triplet upper bound}
                {\bf if} $v > \triangle_w$ {\bf then} $continue$\;
                \nl $w \gets |e_x \cap e_y \cap e_z|$\;
                \nl {\bf if} $\alpha=2$ {\bf then} $w \gets \frac{v - w}{w + 1}$\;
                \nl {\bf if} $w \leq \triangle_w$ {\bf then} continue\;
                \nl {\bf if} $|\triangle_k| < k$ {\bf then} $\triangle \gets \triangle \cup \{e_x, e_y, e_z\}$\;
                \nl \If {$|\triangle_k| = k$}{
                    \nl $\triangle_{min} \gets $ minimum weight triplet in $\triangle_k$\;
                    \nl replace $\triangle_{min}$ in $\triangle_k$ with $\{e_x, e_y, e_z\}$\;
                    \nl $\triangle_w \gets$ minimum weight in $\triangle_k$\;
                }
            }
        }
        \nl \If {$\theta(e_y, \triangle_w, \alpha)$}{ \label{line:m:Sstart}
            \nl \ForEach {$z, N_{yz} \in T$ s.t. $\theta(e_z, \triangle_w, \alpha)$}{
                \nl $S[z][y] \gets N_{yz}$ \tcp*{limit duplicate set ops}
            }
        } \label{line:m:Send}
    }
    \nl \textbf{return} $\triangle_k$
\end{algorithm}

Given a hyperedge triplet $\{e_x, e_y, e_z\}$, let $x = |e_x|, y = |e_y|, z = |e_z|, xy = |e_x \cap e_y|, xz = |e_x \cap e_z|, yz = |e_y \cap e_z|$, and $xyz = |e_x \cap e_y \cap e_z|$.
We can then represent the independent ($W_1$), disjoint ($W_2$), and common weight ($W_3$) formulations in Equation \ref{eq:wght} as:
\begin{equation} \label{eq:ind}
    W_1 = \frac{min(x - xy - xz, y - xy - yz, z - xz - yz) + xyz}{xy + xz + yz - 2 \cdot xyz + 1}
\end{equation}
\begin{equation} \label{eq:dis}
    W_2 = \frac{min(xy, xz, yz) - xyz}{xyz + 1}
\end{equation}
\begin{equation} \label{eq:com}
    W_3 = xyz
\end{equation}

To speed up computation, we can skip over hyperedges based on the upper bounds of their weights.
Unlike the $O(1)$ time it takes to find the sizes of the independent regions $\{x, y, z\}$, the disjoint $\{xy, xz, yz\}$ and common regions $\{xyz\}$ take $O(|V|)$ time for their set intersection operations.
We can achieve speedup by avoiding unnecessary set intersection operations for hyperedge triplets with insufficient weight upper bounds.
This upper bound then becomes tighter as additional region sizes are computed.
Without loss of generality, we have the following inequalities for $\{x, y, z\}$, which give us new upper bounds: $xyz \le min(xy, xz, yz)$ and $\floor{\frac{x}{2}} \ge min(xy/z, xz/y, yz/x)$.
Let $\hat{W_1}$ be equal to the substitution of $min(xy, xz, yz)$ for $xyz$ in $W_1$.
Then we have the following inequalities which provide new upper bounds:

\begin{itemize}
    \item {\bf Independent.} $x \ge \frac{min(x, y) - xy}{xy + 1} \ge \hat{W_1} \ge W_1$
    \item {\bf Disjoint.} $\floor{\frac{x}{2}} \ge W_2, xy \ge min(xy, xz, yz) \ge W_2$
    \item {\bf Common.} $x \ge xy \ge min(xy, xz, yz) \ge W_3$
\end{itemize}

Algorithm \ref{alg:max}, \textsc{Max}, presents our improved framework for the {\sc Disjoint} and {\sc Common} variants---pseudocode for the {\sc Independent} variant is similar. % but longer and moved to the Appendix (Algorithm \ref{alg:max-independent}).
We use the new upper bounds above as part of an ``early stopping'' scheme.
For all of our algorithms, we start by renaming hyperedge ids by decreasing degrees (higher cardinality hyperedges have a lower id) and sorting neighbor lists in ascending order.
%This allows for an "early stopping" procedure where we can easily ignore hyperedges with cardinalities $\le$ to a specific value (all hyperedges with higher ids have a cardinality at most equal to the current hyperedge).
%We can then iterate through all hyperedges (Line \ref{line:mi:b} - Line \ref{line:mi:c}) whose cardinality is $\le$ the current maximum weight.
Then, we iterate through all pairs of hyperedges with at least one common node.
Note that we skip any hyperedge whose upper bound is not greater than the current maximum weight.
%Any hyperedge triplet containing this hyperedge must have a weight of at most its cardinality, and thus these lower cardinality hyperedges can be ignored in the traversal process.
%Since the cardinality of a corresponding hyperedge $C$ may be lower than that of $b$, we also perform this check for all $C$s (Line \ref{line:mi:c}).
Next, we store all non-empty hyperedge pairwise intersections in a local container $T$.
%This initial process is also used in \md\ and \mc.
Afterwards, we iteratively traverse through candidate hyperedge triplets and compute their weight.
Throughout this process, we update the upper bound criteria accordingly and prematurely stop weight computation if the upper bound does not exceed the target weight $\triangle_w$.
Our final step is storing the global set intersections in a container $S$ to limit duplicate set intersection computations.

Our algorithms can be adapted to find hyperedge triplets with a weight above a desired threshold.
We can fix $\triangle_w$ to a set threshold and return all hyperedge triplets whose weight exceeds $\triangle_w$.

{\it Time Complexity.} All three of our variants have the same time complexity. In the worst-case, there exists no maximum hyperedge triplets with a positive weight and hence the time complexity is the same as Algorithm~\ref{alg:basic}, which is $O({|E| \choose 3} \cdot |V|)$. However, due to this being a rare scenario in real-world hypergraphs, this is a very loose bound and our new algorithms are typically much faster than Algorithm\ \ref{alg:basic}, as shown in Section \ref{sec:runtime}.

{\it Space Complexity.} For all three variants of the new algorithm, global container $S$ (Line \ref{line:m:s}) dominates the space complexity of container $T$ (Line \ref{line:m:t}). The space complexity here is the same as Algorithm\ \ref{alg:basic}, which is $O(m \cdot |E|)$.

\subsection{Applications} \label{sec:app}
We consider two higher-level applications for our hyperedge triplets.

\noindent{\bf Local search.} Our algorithms can be adapted to find top-$k$ hyperedge triplets containing a given hyperedge query.
We can iterate over hyperedge triplets containing a queried hyperedge and output those with the highest weights.
This provides a way of exploring maximal hypergraph motifs in the context of a chosen hyperedge of interest.
Local search can also give an insight about the nature of the query hyperedge by comparing the weights of its top-$k$ independent, disjoint, and common triplets.
For example, a hyperedge associated with relatively high common weights compared to its independent weights is typically bundled with other hyperedges.
%\erdem{and even comparing the top weights of three variants can tell about the nature of hpyeredge}
 %We analyzed various datasets with our objective functions in Section \ref{sec:local_subgraphs}.

\noindent{\bf Large subgraphs of triplets.} Although maximal hyperedge triplets are informative, they only describe the relationships between 3 hyperedges.
%However, one can also be interested in subgraphs with more than three hyperedges.
We can obtain subgraphs with more than 3 hyperedges by combining hyperedge triplets into larger clusters of closely-related hyperedges. 
The idea is to merge overlapping hyperedge triplets with high weights to find groups of similar hyperedges.
%To do so, we can collect multiple hyperedge triplets with a significant common weight.
%As stated in Section~\ref{sec:maxAlgs}, we can adapt our algorithms to store all hyperedge triplets whose weight is above a certain threshold.
%After this hyperedge triplet collection process, we then create an edge-weighted graph where nodes represent hyperedges, edges connect nodes which are in a hyperedge triplet, and edge weights reflect the number of shared triplets to which that pair of hyperedges belong. 
We can create an edge-weighted graph where nodes represent hyperedges, edges connect nodes which are in a hyperedge triplet, and edge weights reflect the number of shared triplets to which that pair of hyperedges belong. 
The connected components of this graph then represent larger groups of hyperedges that are closely related. % represent the desired subgraphs.

\section{Results} \label{sec:results}

\begin{table}[!t]
\centering
\caption{Dataset details. $|V|$ is the number of nodes, $|E|$ is the number of hyperedges, $m$ is the sum of node degrees (or hyperedge sizes), and $d_E^*$ is the maximum hyperedge size.}
\label{tab:datasets}
\begin{tabular}{|l|r|r|r|r|} 
 \hline
 \multicolumn{1}{|c|}{Dataset} & \multicolumn{1}{c|}{$|V|$} & \multicolumn{1}{c|}{$|E|$} & \multicolumn{1}{c|}{$m$} & \multicolumn{1}{c|}{$d_E^*$}\\
 \hline
 \GDx & 17,549 & 24,444 & 628,566 & 5,053\\
 \FOx & 25,076 & 178,265 & 718,379 & 1,091\\
 \DBx & 258,769 & 7,783 & 463,497 & 24,821\\
 %\MLx & 283,228 & 53,888 & 27,753,442 & 97,999\\
 \YEx & 1,987,929 & 150,346 & 6,745,760 & 7,568\\
 \ATx & 1,775,852 & 559,775 & 5,727,435 & 5,730\\
 \ABx & 15,362,619 & 2,930,451 & 51,062,224 & 58,147\\
 \hline
\end{tabular}
\end{table}

In this section, we evaluate our algorithms on several real-world hypergraphs.
%In Section \ref{sec:runtime}, we provide our runtime results which compare the runtimes of our algorithms.
%Then, in Section \ref{sec:exp}, we show the top 10 weights for each dataset.
We first explore the distribution of nodes in our maximum hyperedge triplets through a study on entropy in Section \ref{sec:entropy}.
Then, we present an in-depth case study on the \YE\ dataset in Section~\ref{sec:ye}.
Next, in Sections~\ref{sec:local_subgraphs}\ and~\ref{sec:large_subgraphs}, we discuss the results of the applications built on hyperedge triplets, as discussed in Section~\ref{sec:app}.
Lastly, we give the runtime results in Section~\ref{sec:runtime}.
%Lastly, in Section~\ref{sec:domain}, we compare our top-$k$ weight distributions with $h$-motif characteristic profiles for the task of graph classification.

Table \ref{tab:datasets}\ presents the statistics of our real-world hypergraphs used in experiments up to Section~\ref{sec:runtime}.
\GD~(\GDx) is a network of gene-disease associations from DisGeNET~\cite{pinero2016disgenet}.
\FO~(\FOx) contains recipe reviews covering 18 years of user interactions on food.com~\cite{majumder2019generating}.
\DB~(\DBx) connects artistic works or artists with genres from DBpedia~\cite{auer2007dbpedia}.
%\ML~(\MLx) contains movie ratings for users of MovieLens~\cite{harper2015movielens}.
\YE~(\YEx) features user business reviews from Yelp~\cite{yelp}.
\AT~(\ATx) and \AB~(\ABx) contain user ratings on the tools/home improvement and books categories of Amazon, respectively~\cite{ni2019justifying}.
%\jason{Not sure what use case you are referencing; perhaps the amazon use case? Two hyperedges (A,B) may be in the same hyperedge triplet (ABC, ABD), which is what it was referencing (I changed wording to "common hyperedges" instead).}
%\GDER~(\GDERx) and \GDCL~(\GDCLx) are two synthetic hypergraphs based on \GD\ generated by the Erd\H{o}s-R\'enyi and Chung-Lu models \cite{aksoy2017measuring}.

All experiments are performed on a Linux operating system with an Intel Xeon Gold processor at 2.1 GHz and 256GB memory.
%The processor contains 2 sockets with each having 20 cores for a total of 40 cores.
We implemented our algorithms in C++ and compiled using GCC 7.1.0 at the -O3 level.
Our code is available at https://anonymous.4open.science/r/hyperedge-triplets-blinded.

\subsection{Node Distributions with Entropy} \label{sec:entropy}

\begin{figure}[!t]
    \centering
    \begin{subfigure}[h]{0.49\linewidth}
        \centering
        \includegraphics[width=\textwidth]{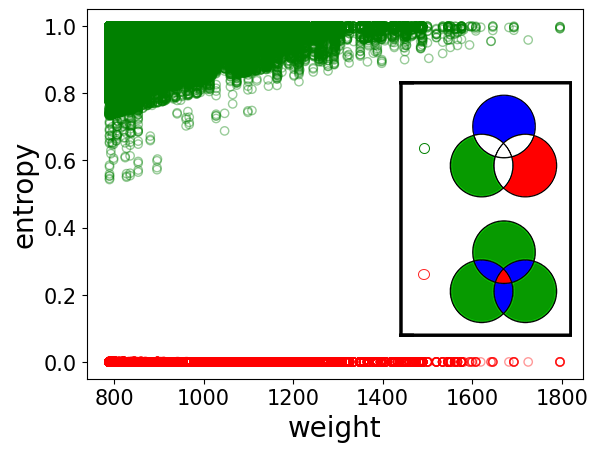}
        \caption{Independent: \YE}
        \label{fig:YE_entropy}
    \end{subfigure}
    \begin{subfigure}[h]{0.49\linewidth}
        \centering
        \includegraphics[width=\textwidth]{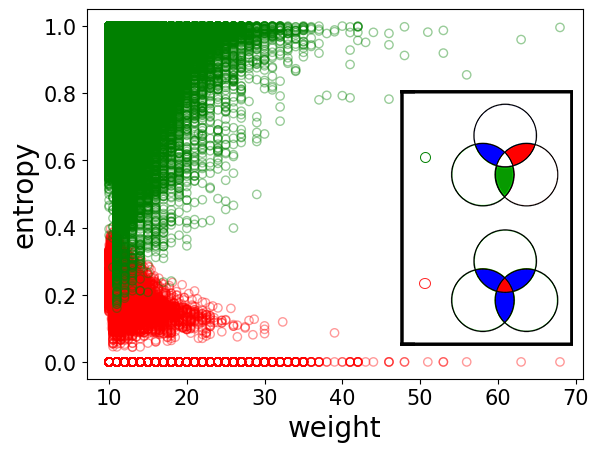}
        \caption{Disjoint: \YE}
        \label{fig:YE_entropy_dis}
    \end{subfigure}
   \caption{Relative entropies of the independent and disjoint regions for the top 1M independent and disjoint triplets, respectively. Entropies are between the colored regions in the legend.}
   \label{fig:entropy}
\end{figure}

We first evaluate the goodness of our weight formulation (Equation~\ref{eq:wght}) from an information theoretical perspective in Figure~\ref{fig:entropy}.
We consider the relative entropies (i.e. Shannon entropies) of the top 1M hyperedge triplets w.r.t. the independent and disjoint variants in \YE.
Relative entropy is defined as $-\displaystyle\sum_{x \in X}^{} p_x \cdot \log(p_x)$ where $p_x$ is the probability of the event $x \in X$.
%Figure~\ref{fig:entropy} shows the results.
For the independent variant (Figure~\ref{fig:YE_entropy}), we compute two entropy values for each triplet: (1) entropy of the three independent regions ($R_1$s in Figure~\ref{fig:regionLabels}), denoted by green circles, and (2) entropy of the sum of independent regions (sum of $R_1$s), sum of disjoint regions (sum of $R_2$s), and triple of the common region (3*$R_3$), denoted by red circles.
The green circles, which are close to 1, imply that nodes tend to be evenly distributed across the independent regions whereas the red circles, which are close to 0, suggest that almost all nodes are in the independent regions.
%Hyperedge triplets with larger independent weights exhibit a higher disparity w.r.t. the independent regions, as denoted by green circles with high entropy values, and low disparity w.r.t. regions of different types ($R_1$, $R_2$, $R_3$), as shown by red circles with low entropy values.
%This shows that the numerator of Equation~\ref{eq:wght} is effective in ensuring the disparity among independent regions and that the normalization w.r.t. the inner regions enable standardization.
This shows that the numerator of Equation~\ref{eq:wght} is effective at establishing parity among independent regions and that the denominator ensures disparity among all region types.
Similarly, for the disjoint variant (Figures~\ref{fig:YE_entropy_dis}), we compute two entropy values for each triplet: (1) entropy of the three disjoint regions ($R_2$s in Figure~\ref{fig:regionLabels}), denoted by green circles, and (2) entropy of the sum of disjoint regions (sum of $R_2$s) and triple of the common region (3*$R_3$), denoted by red circles.
%Again, triplets with larger disjoint weights have a higher disparity w.r.t. the disjoint regions and low disparity w.r.t. regions of different types.
Again, triplets with larger disjoint weights have higher parity w.r.t. the disjoint regions and higher disparity w.r.t. regions of different types.
%Our problem formulations can thus be viewed through an information theoretical perspective: we target hyperedge triplets which maximize the disparity among considered regions.
Our problem formulations can thus be viewed through an information theoretical perspective: we target hyperedge triplets which maximize the disparity among region types.
% That is, we seek hyperedge triplets where a significant proportion of nodes are evenly distributed across the corresponding region type.

\subsection{Case Study: \YE} \label{sec:ye}
In this section, we present the top hyperedge triplets for each objective---independent, disjoint, and common---on~\YE~\cite{yelp}.

\begin{figure}[!t]
\centering
\includegraphics[width=\linewidth]{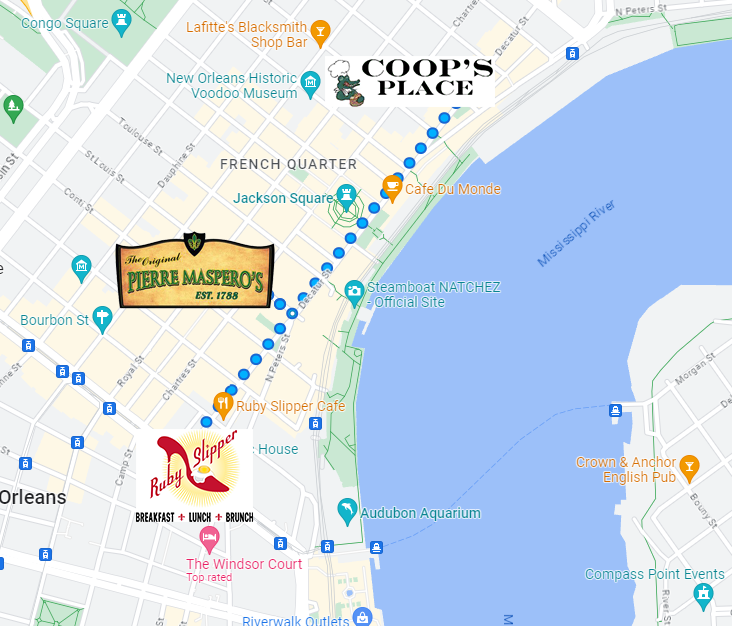}
\caption{The maximum disjoint hyperedge triplet in \YE: {\it Ruby Slipper Cafe}, {\it Coop's Place}, and {\it The Original Pierre Maspero's} in New Orleans.}
\label{fig:ye_md}
\end{figure}

%\begin{itemize}
\noindent {\bf Independent.} The maximum hyperedge triplet has {\it The Eagle}, {\it Palace Caf\'e}, and {\it Mesa Verde} restaurants with a weight of 1796.
All three are highly reviewed restaurants in different cities across the United States (Indianapolis with 2233 reviews, New Orleans with 1822 reviews, and Santa Barbara with 1796 reviews, respectively) and serve different cuisines.

\noindent  {\bf Disjoint.} The maximum hyperedge triplet contains three restaurants in New Orleans with a weight of 68.
{\it Ruby Slipper Cafe} (A), {\it Coop's Place} (B), and {\it The Original Pierre Maspero's} (C) are all within walking distance, as shown in Figure~\ref{fig:ye_md}.
{\it Ruby Slipper Cafe} is a breakfast and brunch restaurant while others serve Cajun/Creole food with the {\it Coop's Place} being a bar and the {\it The Original Pierre Maspero's} being a sit-down restaurant.
Interestingly, although many users have reviewed two of the three restaurants, no users have reviewed all three.
Checking the keywords in the reviews, we observe a division of users who go out to eat breakfast with around 35\% of the reviews mentioning breakfast for $AB$ and $AC$ but only about 6\% for $BC$, where lunch and dinner are more common.
With an approximately 25\% higher proportion of reviews describing the bar experience for $AB$ compared to $AC$, the difference between $AB$ and $AC$ seems to rely upon the preference for bars or restaurants.

From these findings, the management of $A$ now know that many of its interested customers often go to $B$ or $C$ based on their bar or traditional restaurant preference, respectively. If $B$ is closed for a certain period of time, the management of $A$ can feature its bar offerings to improve business during this period. As such, hyperedge triplets are an effective building block for sites such as $Yelp$ to help businesses make more informed decisions.\\

\noindent {\bf Common.} With a weight of 244, {\it Reading Terminal Market}, {\it Pat's King of Steaks}, and {\it Geno's Steaks} are three businesses in Philadelphia within 2 miles from each other with 5,721, 4,250, and 3,401 total reviews, respectively.
{\it Reading Terminal Market} is a popular farmers market while {\it Pat's King of Steaks} and {\it Geno's Steaks} are cheesesteak stalls right next to each other.
The rivalry between Pat's and Geno's cheesesteaks has sparked a well-known, contested debate.
Along with the historic tourist destination of {\it Reading Terminal Market}, three of Philadelphia's most popular tourist spots are captured here thanks to the users who predominantly review all three places.
%\end{itemize}
%

\subsection {Local Hyperedge Triplets} \label{sec:local_subgraphs}
%Our algorithms easily adapt to find maximum hyperedge triplets containing a specified hyperedge.
%We simply iterate over all hyperedge triplets containing a user-inputted hyperedge, outputting the one with the highest weight. This provides a way of exploring maximal hypergraph motifs in the context of a chosen hyperedge of interest. We analyzed various datasets with our objective functions.\\
Here, we provide a case study for the local search around a query hyperedge, described in Section~\ref{sec:app}. We select the {\it Really Good Vegetarian Meatloaf (Really!)} recipe, a niche vegetarian meal, in the \FO~dataset and find the top-10 maximum triplets for all three variants around it.

\noindent {\bf Independent.} All of the top-10 independent triplets (best weight: 128) contain a meat option, such as "fabulous beef stew" or "tortured chicken - beer can", reflecting an opposite taste for vegetarians. The third option is typically a healthy meat option such as "zesty low fat chicken breasts" or "potato salad with chipotle peppers (a man's salad)" which contains bacon. Note that the recipes with meat options also offer distinct features of being healthy or not.\\ 
\noindent {\bf Disjoint.} The top-10 disjoint triplets (best weight: 12) often include a dessert like "thick chocolate pudding" and "oatmeal cottage cheese pancakes" along with a chicken meal such as "creamy cajun chicken pasta" and "amazing chicken marinade". Note that the top disjoint weight (12) is much smaller than the top independent weight (128).\\
\noindent {\bf Common.} The top common weight (9) is even lower than disjoint's, suggesting that there is not much correlation with other recipes.

\subsection {Larger Patterns Via Triplet Merging} \label{sec:large_subgraphs}
%Although maximal hyperedge triplets are informative, they only %describe the relationships between three hyperedges.
%However, we may also be interested in relationships identified by increasing the number of hyperedges within each maximum motif. One way to do this is by merging hyperedge triplets into larger clusters of closely-related hyperedges. Here, we briefly introduce a possible approach towards this end, which we tested on our largest network, \AB.

%The main idea is to merge overlapping hyperedge triplets with high weights to find groups of similar hyperedges.
%To do so, we need to collect multiple hyperedge triplets with a significant common weight.
%As stated in Section \ref{sec:maxAlgs}, we can adapt Algorithm \ref{alg:max}\ to store all hyperedge triplets whose weight is above a certain threshold.
%After this hyperedge triplet collection process, we then create an edge-weighted graph where nodes represent hyperedges, edges connect nodes which are in a hyperedge triplet, and edge weights reflect the number of shared triplets to which that pair of hyperedges belong. 
%The connected components of this graph identify larger groups of hyperedges that are closely related via common weight. % represent the desired subgraphs.

\begin{figure}[t]
    \centering
    \includegraphics[width=0.4\textwidth]{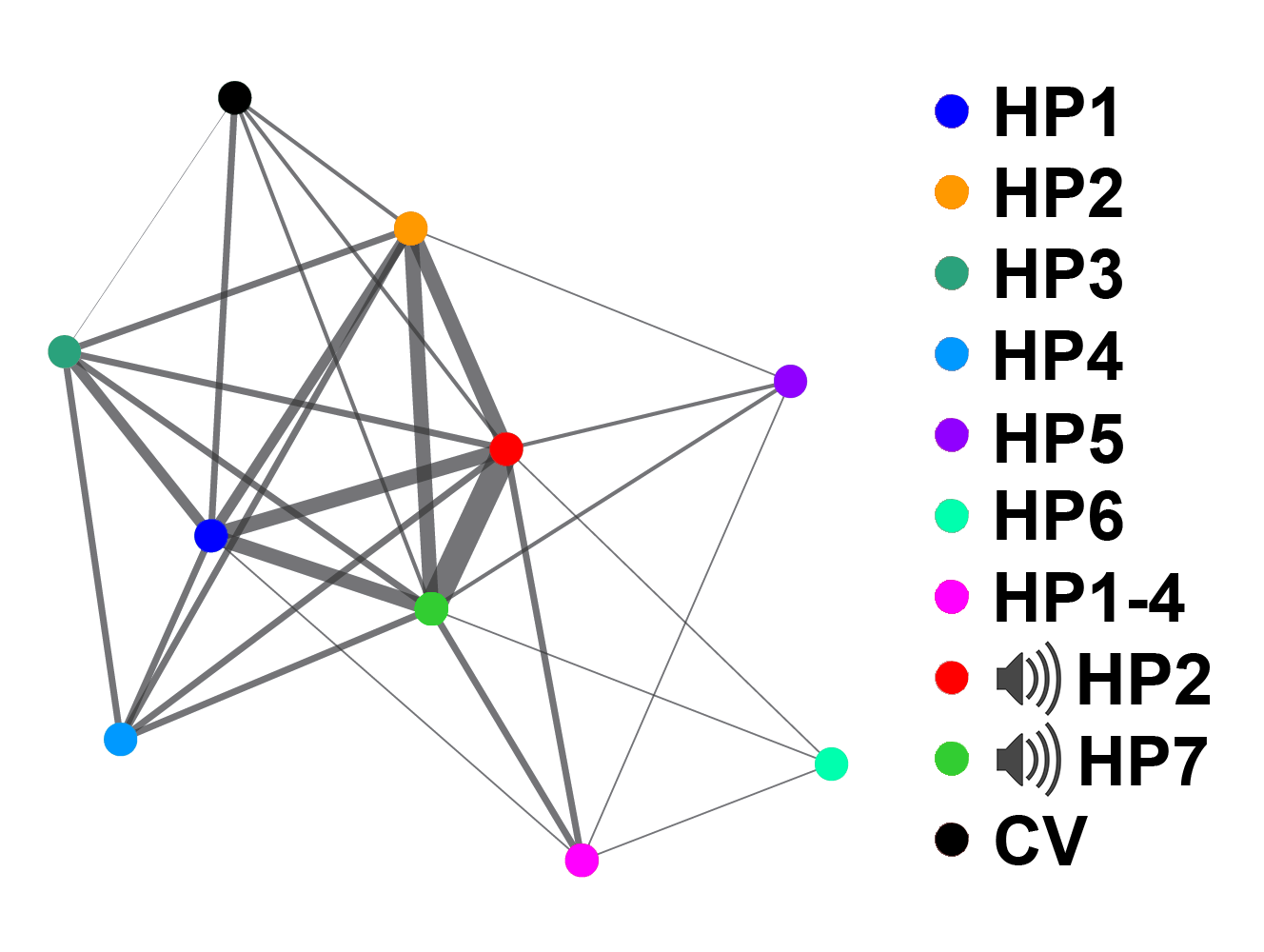}
    \caption{\AB's largest connected component for common weights of at least 500. The edge thickness in the visualization is proportional to the number of triplets containing the connecting book products. There are 6 Harry Potter books, one Harry Potter box set, 2 Harry Potter audiobooks, and a Casual Vacancy novel.}
    \label{fig:ab_large_common}
\end{figure}

We now consider the hyperedge triplet merging approach described in Section~\ref{sec:app} on \AB\ using hyperedge triplets with a common weight of at least 500.
The largest connected component in the results consists of ten J.K. Rowling novels with nine related to the {\it Harry Potter} series and one {\it The Casual Vacancy} edition.
Figure~\ref{fig:ab_large_common} shows the graph of this connected component with triangles representing hyperedge triplets and edge thickness denoting the number of triplets containing the book pair.
The first two Harry Potter novels along with the two audiobooks participate in many triplets and form a central cluster.
This follows the trend that the first books in a series are typically more popular than later books and that many readers prefer to own both a physical copy and the corresponding audiobook.
In regards to multiple physical copies, readers typically do not purchase and review multiple physical copies of the same book, as shown with the Harry Potter box set lacking significant connectivity with its individual novels.

\subsection{Runtime Experiments} \label{sec:runtime}

\begin{table}[!t]
\centering
\caption{Runtime results (in seconds). "$-$" denotes timed-out runs that took more than 24 hours.}
\label{tab:runtime}
\begin{tabular}{|l|r|r|r|r|r|r|} 
 \hline
 \multicolumn{1}{|c|}{} & \multicolumn{2}{c|}{\sc Independent} & \multicolumn{2}{c|}{\sc Disjoint} & \multicolumn{2}{c|}{\sc Common}\\
 \cline{2-7}
 \multicolumn{1}{|c|}{Net.} & \multicolumn{1}{c|}{\textsc{Basic}} & \multicolumn{1}{c|}{\textsc{Max}} & \multicolumn{1}{c|}{\textsc{Basic}} & \multicolumn{1}{c|}{\textsc{Max}} & \multicolumn{1}{c|}{\textsc{Basic}} & \multicolumn{1}{c|}{\textsc{Max}}\\
 \hline
 \GDx & 4.72K & 18.78 & 452.28 & 5.97 & 15.30 & 0.18\\
 \DBx & 14.39 & 0.24 & 0.32 & 0.24 & 0.30 & 0.22\\
 \YEx & 20.51K & 2.93 & 2.46K & 4.82 & 1.18K & 3.41\\
 \ATx & - & 2.72 & 145.94 & 3.41 & 43.47 & 2.86\\
 \ABx & - & 32.61 & 68.75K & 42.58 & 1.93K & 24.19\\
 \hline
\end{tabular}
\end{table}

\begin{figure}[t]
    \caption{Runtimes (left) and number of processed triplets (right) for \textsc{Max} when varying $k$ on \ABx.}
    \vspace{-1ex}
    \label{fig:ab_runtimek}
    \centering
    \begin{subfigure}[h]{0.48\linewidth}
        \centering
        \includegraphics[width=\textwidth]{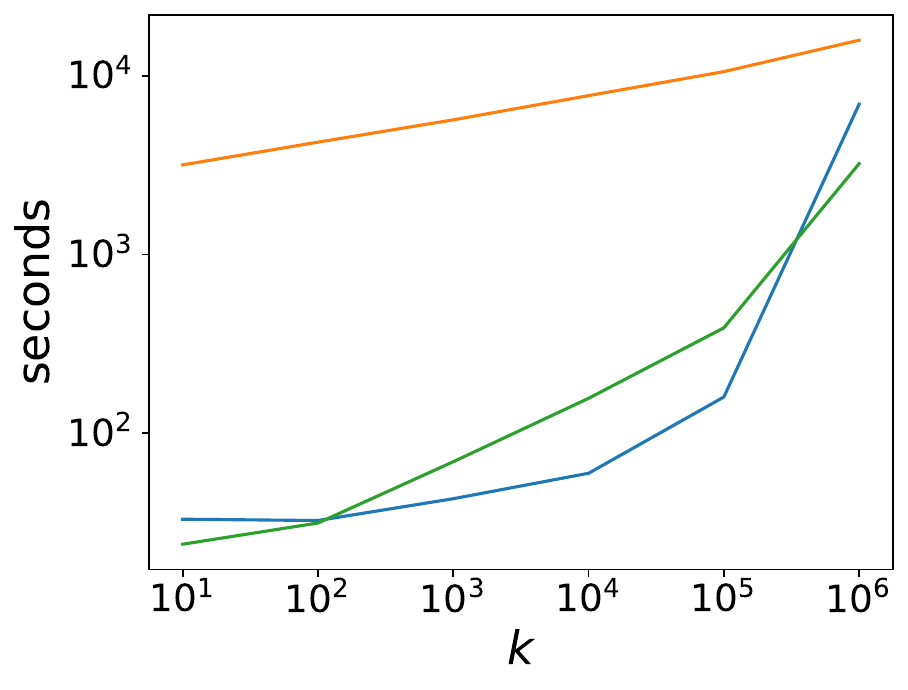}
    \end{subfigure}
    \begin{subfigure}[h]{0.48\linewidth}
        \centering
        \includegraphics[width=\textwidth]{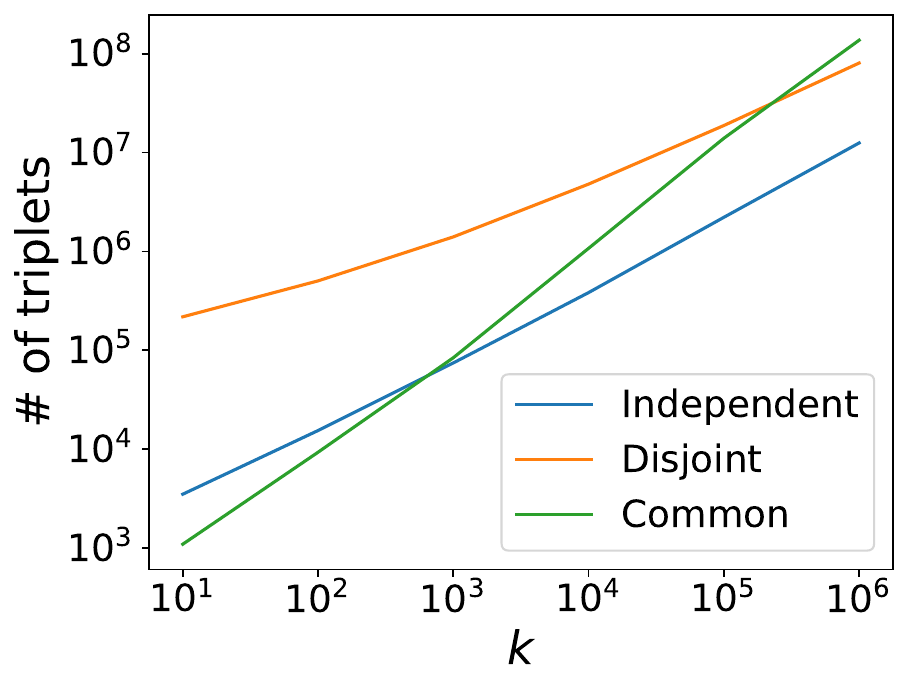}
    \end{subfigure}
\end{figure}

We compare our improved algorithms against the naive baseline for finding the top-1 triplet for each variant.
Table~\ref{tab:runtime} gives the results.
\bi\ is not able to finish computation in under 24 hours for \ATx\ and \ABx.
%Algorithm \ref{alg:max}\ variants (\mi, \md, and \mc) significantly outperform Algorithm \ref{alg:basic}\ variants (\bi, \bd, and \bc).
\mi, \md, and \mc\ significantly outperform Algorithm \ref{alg:basic}\ variants (\bi, \bd, and \bc).
Speedups vary between 1.3x and 7,000x.

When considering the top-$k$ triplets, Figure \ref{fig:ab_runtimek} shows the runtimes and number of processed triplets for our largest network, \ABx, when varying $k$.
Our best algorithms are able to find and rank the top 1 million hyperedge triplets in about 1.9, 4.4, and 0.9 hours for our independent, disjoint, and common formulations, respectively. 
We achieve this by processing only a small proportion of all $|E|^3$ hyperedge triplets.
Real-world hypergraphs typically have skewed degree distributions where there are a few hyperedges with very large cardinality and many with low cardinality.
It is expected that hyperedges of larger cardinalities tend to participate in higher weight triplets.
As a result, our ``early stopping'' considerations in Algorithm~\ref{alg:max} have a significant impact on the runtime even though its time complexity is the same as its corresponding \textsc{Basic} variants.

\section{Conclusion and Future Work} \label{sec:conclusion}
In this work, we introduced a novel problem formulation for finding different types of top-$k$ hyperedge triplets. The connection patterns we formulate take advantage of the quality offered by the intersection sizes in the ranking and discovery of hypergraph motifs.
We proposed efficient algorithms to materialize each of those formulations in real-world data.
%Through case studies on the \YE\ and \AB\ datasets, we give several interesting categorical relationships not previously found in real-world hypergraphs.
%Through case studies on the \YE\ and \AB\ datasets, we can estimate the proximity of businesses without the use of any metadata and the existence of behavioral trends among the readers of the most popular Amazon books, respectively.
Experiments and case studies on real-world hypergraphs show practical benefit and efficient computation on networks with millions of nodes and edges.

Although $h$-motifs are designed to classify (small) hypergraphs, they are trivial for large hypergraphs, unlike hyperedge triplets.
It is well known that many graph neural network architectures also have difficulty classifying large hypergraphs due to their memory and runtime constraints.
Hyperedge triplets provide a building block for fast and memory-efficient large hypergraph classification models, which we will address as future work.
Despite the strong performance exhibited by our algorithms, we are yet to process hypergraphs with billions of edges in a reasonable amount of time, which is another interesting future direction to explore.
Hyperedge triplets can also be used to study hyperedge prediction as the hyperedges with the strongest ties (or weights) are more likely to involve nodes in new hyperedges.

%Although our algorithms exhibit strong performance, we are yet to process hypergraphs with billions of edges in a reasonable amount of time, which is one interesting future direction to explore.
%As a future work, we will investigate further enhancements to our algorithms and parallelization approaches which speedup the computation. 
%Another interesting problem to study is hyperedge prediction by using our novel formulations, as the hyperedges with the strongest ties (or weights) are more likely to involve nodes in new hyperedges.
%
%Further research is also needed on a hyperedge triplet merging scheme, with our initial approach having promising results.
%An interesting avenue is to use a merging scheme to generate a regional map of business locations in \YE\ using only the graph data.

\section{Acknowledgments}
    \noindent This research was supported by NSF award OAC-2107089. PNNL
    Information Release: PNNL-SA-191147. Pacific Northwest National
    Laboratory is operated by Battelle Memorial Institute under Contract DE-ACO6-76RL01830.

\bibliographystyle{IEEEtran}
\bibliography{hyper}

% Generated by IEEEtran.bst, version: 1.12 (2007/01/11)
\begin{thebibliography}{10}
\providecommand{\url}[1]{#1}
\csname url@samestyle\endcsname
\providecommand{\newblock}{\relax}
\providecommand{\bibinfo}[2]{#2}
\providecommand{\BIBentrySTDinterwordspacing}{\spaceskip=0pt\relax}
\providecommand{\BIBentryALTinterwordstretchfactor}{4}
\providecommand{\BIBentryALTinterwordspacing}{\spaceskip=\fontdimen2\font plus
\BIBentryALTinterwordstretchfactor\fontdimen3\font minus
  \fontdimen4\font\relax}
\providecommand{\BIBforeignlanguage}[2]{{%
\expandafter\ifx\csname l@#1\endcsname\relax
\typeout{** WARNING: IEEEtran.bst: No hyphenation pattern has been}%
\typeout{** loaded for the language `#1'. Using the pattern for}%
\typeout{** the default language instead.}%
\else
\language=\csname l@#1\endcsname
\fi
#2}}
\providecommand{\BIBdecl}{\relax}
\BIBdecl

\bibitem{zhou2006learning}
D.~Zhou, J.~Huang, and B.~Sch{\"o}lkopf, ``Learning with hypergraphs:
  Clustering, classification, and embedding,'' \emph{Advances in neural
  information processing systems}, vol.~19, 2006.

\bibitem{newman2001scientific}
M.~E. Newman, ``Scientific collaboration networks. i. network construction and
  fundamental results,'' \emph{Physical Review E}, vol.~64, no.~1, p. 016131,
  2001.

\bibitem{robins2004small}
G.~Robins and M.~Alexander, ``Small worlds among interlocking directors:
  Network structure and distance in bipartite graphs,'' \emph{Computational \&
  Mathematical Organization Theory}, vol.~10, no.~1, pp. 69--94, 2004.

\bibitem{watts1998collective}
D.~J. Watts and S.~H. Strogatz, ``Collective dynamics of
  ‘small-world’networks,'' \emph{Nature}, vol. 393, no. 6684, pp. 440--442,
  1998.

\bibitem{agarwal2006higher}
P.~Agarwal, ``Higher education in india: The need for change,'' Working paper,
  Tech. Rep., 2006.

\bibitem{hayashi2020hypergraph}
K.~Hayashi, S.~G. Aksoy, C.~H. Park, and H.~Park, ``Hypergraph random walks,
  laplacians, and clustering,'' in \emph{Proceedings of the 29th acm
  international conference on information \& knowledge management}, 2020, pp.
  495--504.

\bibitem{coletto2017motif}
M.~Coletto, K.~Garimella, A.~Gionis, and C.~Lucchese, ``A motif-based approach
  for identifying controversy,'' in \emph{Eleventh International AAAI
  Conference on Web and Social Media}, 2017.

\bibitem{ma2014motif}
W.~Ma, W.~S. Noble, and T.~L. Bailey, ``Motif-based analysis of large
  nucleotide data sets using meme-chip,'' \emph{Nature Protocols}, vol.~9,
  no.~6, pp. 1428--1450, 2014.

\bibitem{sariyuce2018peeling}
A.~E. Sar{\i}y{\"u}ce and A.~Pinar, ``Peeling bipartite networks for dense
  subgraph discovery,'' in \emph{Proceedings of the Eleventh ACM International
  Conference on Web Search and Data Mining}, 2018, pp. 504--512.

\bibitem{lee2020hypergraph}
G.~Lee, J.~Ko, and K.~Shin, ``Hypergraph motifs: Concepts, algorithms, and
  discoveries,'' \emph{Proceedings of the VLDB Endowment}, vol.~13, no.~11, pp.
  2256--2269, 2020.

\bibitem{aksoy2020hypernetwork}
S.~G. Aksoy, C.~Joslyn, C.~O. Marrero, B.~Praggastis, and E.~Purvine,
  ``Hypernetwork science via high-order hypergraph walks,'' \emph{EPJ Data
  Science}, vol.~9, no.~1, p.~16, 2020.

\bibitem{kumar2020retrieving}
R.~Kumar, P.~Liu, M.~Charikar, and A.~R. Benson, ``Retrieving top weighted
  triangles in graphs,'' in \emph{Proceedings of the 13th International
  Conference on Web Search and Data Mining}, 2020, pp. 295--303.

\bibitem{taniguchi2022efficient}
R.~Taniguchi, D.~Amagata, and T.~Hara, ``Efficient retrieval of top-k weighted
  spatial triangles,'' in \emph{International Conference on Database Systems
  for Advanced Applications}.\hskip 1em plus 0.5em minus 0.4em\relax Springer,
  2022, pp. 224--231.

\bibitem{taniguchi2022efficient2}
------, ``Efficient retrieval of top-k weighted triangles on static and dynamic
  spatial data,'' \emph{IEEE Access}, vol.~10, pp. 55\,298--55\,307, 2022.

\bibitem{zhang2023top}
F.~Zhang, X.~Gou, and L.~Zou, ``Top-k heavy weight triangles listing on graph
  stream,'' \emph{World Wide Web}, vol.~26, no.~4, pp. 1827--1851, 2023.

\bibitem{kosyfaki2019flow}
C.~Kosyfaki, N.~Mamoulis, E.~Pitoura, and P.~Tsaparas, ``Flow motifs in
  interaction networks,'' Ph.D. dissertation, University of Ioannina, 2019.

\bibitem{wang2014rectangle}
J.~Wang, A.~W.-C. Fu, and J.~Cheng, ``Rectangle counting in large bipartite
  graphs,'' in \emph{2014 IEEE International Congress on Big Data}.\hskip 1em
  plus 0.5em minus 0.4em\relax IEEE, 2014, pp. 17--24.

\bibitem{sanei2018butterfly}
S.-V. Sanei-Mehri, A.~E. Sariyuce, and S.~Tirthapura, ``Butterfly counting in
  bipartite networks,'' in \emph{Proceedings of the 24th ACM SIGKDD
  International Conference on Knowledge Discovery \& Data Mining}, 2018, pp.
  2150--2159.

\bibitem{chiba1985arboricity}
N.~Chiba and T.~Nishizeki, ``Arboricity and subgraph listing algorithms,''
  \emph{SIAM Journal on computing}, vol.~14, no.~1, pp. 210--223, 1985.

\bibitem{sanei2019fleet}
S.-V. Sanei-Mehri, Y.~Zhang, A.~E. Sariy{\"u}ce, and S.~Tirthapura, ``Fleet:
  Butterfly estimation from a bipartite graph stream,'' in \emph{Proceedings of
  the 28th ACM International Conference on Information and Knowledge
  Management}, 2019, pp. 1201--1210.

\bibitem{shi2022parallel}
J.~Shi and J.~Shun, ``Parallel algorithms for butterfly computations,'' in
  \emph{Massive Graph Analytics}.\hskip 1em plus 0.5em minus 0.4em\relax
  Chapman and Hall/CRC, 2022, pp. 287--330.

\bibitem{lind2005cycles}
P.~G. Lind, M.~C. Gonz{\'a}lez, and H.~J. Herrmann, ``Cycles and clustering in
  bipartite networks,'' \emph{Physical Review E}, vol.~72, no.~5, p. 056127,
  2005.

\bibitem{aksoy2017measuring}
S.~G. Aksoy, T.~G. Kolda, and A.~Pinar, ``Measuring and modeling bipartite
  graphs with community structure,'' \emph{Journal of Complex Networks},
  vol.~5, no.~4, pp. 581--603, 2017.

\bibitem{yang2021efficient}
Y.~Yang, Y.~Fang, M.~E. Orlowska, W.~Zhang, and X.~Lin, ``Efficient bi-triangle
  counting for large bipartite networks,'' \emph{Proceedings of the VLDB
  Endowment}, vol.~14, no.~6, pp. 984--996, 2021.

\bibitem{schank2005approximating}
T.~Schank and D.~Wagner, ``Approximating clustering coefficient and
  transitivity.'' \emph{Journal of Graph Algorithms and Applications}, vol.~9,
  no.~2, pp. 265--275, 2005.

\bibitem{opsahl2013triadic}
T.~Opsahl, ``Triadic closure in two-mode networks: Redefining the global and
  local clustering coefficients,'' \emph{Social Networks}, vol.~35, no.~2, pp.
  159--167, 2013.

\bibitem{niu2022counting}
J.~Niu, J.~Zola, and A.~E. Sar{\i}y{\"u}ce, ``Counting induced 6-cycles in
  bipartite graphs,'' in \emph{Proceedings of the 51st International Conference
  on Parallel Processing}, 2022, pp. 1--10.

\bibitem{lotito2022higher}
Q.~F. Lotito, F.~Musciotto, A.~Montresor, and F.~Battiston, ``Higher-order
  motif analysis in hypergraphs,'' \emph{Communications Physics}, vol.~5,
  no.~1, p.~79, 2022.

\bibitem{lotito2023exact}
Q.~F. Lotito, F.~Musciotto, F.~Battiston, and A.~Montresor, ``Exact and
  sampling methods for mining higher-order motifs in large hypergraphs,''
  \emph{Computing}, pp. 1--20, 2023.

\bibitem{Chodrow20}
P.~S. Chodrow, ``Configuration models of random hypergraphs,'' \emph{Journal of
  Complex Networks}, vol.~8, no.~3, p. cnaa018, 2020.

\bibitem{lee2024hypergraph}
G.~Lee, S.~Yoon, J.~Ko, H.~Kim, and K.~Shin, ``Hypergraph motifs and their
  extensions beyond binary,'' \emph{The VLDB Journal}, vol.~33, no.~3, pp.
  625--665, 2024.

\bibitem{landry2023filtering}
N.~W. Landry, I.~Amburg, M.~Shi, and S.~G. Aksoy, ``Filtering higher-order
  datasets,'' \emph{arXiv preprint arXiv:2305.06910}, 2023.

\bibitem{pinero2016disgenet}
J.~Pi{\~n}ero, {\`A}.~Bravo, N.~Queralt-Rosinach,
  A.~Guti{\'e}rrez-Sacrist{\'a}n, J.~Deu-Pons, E.~Centeno,
  J.~Garc{\'\i}a-Garc{\'\i}a, F.~Sanz, and L.~I. Furlong, ``Disgenet: a
  comprehensive platform integrating information on human disease-associated
  genes and variants,'' \emph{Nucleic acids research}, p. gkw943, 2016.

\bibitem{miller2011efficient}
J.~C. Miller and A.~Hagberg, ``Efficient generation of networks with given
  expected degrees,'' in \emph{International Workshop on Algorithms and Models
  for the Web-Graph}.\hskip 1em plus 0.5em minus 0.4em\relax Springer, 2011,
  pp. 115--126.

\bibitem{czumaj2009finding}
A.~Czumaj and A.~Lingas, ``Finding a heaviest vertex-weighted triangle is not
  harder than matrix multiplication,'' \emph{SIAM Journal on Computing},
  vol.~39, no.~2, pp. 431--444, 2009.

\bibitem{benson2018simplicial}
A.~R. Benson, R.~Abebe, M.~T. Schaub, A.~Jadbabaie, and J.~Kleinberg,
  ``Simplicial closure and higher-order link prediction,'' \emph{Proceedings of
  the National Academy of Sciences}, vol. 115, no.~48, pp. E11\,221--E11\,230,
  2018.

\bibitem{majumder2019generating}
B.~P. Majumder, S.~Li, J.~Ni, and J.~McAuley, ``Generating personalized recipes
  from historical user preferences,'' in \emph{Proceedings of the 2019
  Conference on Empirical Methods in Natural Language Processing and the 9th
  International Joint Conference on Natural Language Processing
  (EMNLP-IJCNLP)}.\hskip 1em plus 0.5em minus 0.4em\relax Association for
  Computational Linguistics, 2019, pp. 5976--5982.

\bibitem{auer2007dbpedia}
S.~Auer, C.~Bizer, G.~Kobilarov, J.~Lehmann, R.~Cyganiak, and Z.~Ives,
  ``Dbpedia: A nucleus for a web of open data,'' in \emph{The semantic
  web}.\hskip 1em plus 0.5em minus 0.4em\relax Springer, 2007, pp. 722--735.

\bibitem{yelp}
``{Yelp Open Dataset},'' 2023, \url{https://www.yelp.com/dataset/}.

\bibitem{ni2019justifying}
J.~Ni, J.~Li, and J.~McAuley, ``Justifying recommendations using
  distantly-labeled reviews and fine-grained aspects,'' in \emph{Proceedings of
  the 2019 conference on empirical methods in natural language processing and
  the 9th international joint conference on natural language processing
  (EMNLP-IJCNLP)}, 2019, pp. 188--197.

\end{thebibliography}

\end{document}